\documentclass{article}

\usepackage[final]{neurips_2025}

\usepackage[utf8]{inputenc} % allow utf-8 input
\usepackage[T1]{fontenc}    % use 8-bit T1 fonts
\usepackage{hyperref}       % hyperlinks
\usepackage{url}            % simple URL typesetting
\usepackage{booktabs}       % professional-quality tables
\usepackage{amsfonts}       % blackboard math symbols
\usepackage{nicefrac}       % compact symbols for 1/2, etc.
\usepackage{microtype}      % microtypography
\usepackage{xcolor}         % colors
\usepackage{colortbl}
\usepackage{multirow}
\usepackage{amsmath} 
\usepackage{graphicx}
\usepackage{subcaption}
\usepackage{pifont} % Access to PostScript standard Symbol and Dingbats fonts
\usepackage{makecell}

\newcommand{\cellbest}{\cellcolor{red!40}}
\newcommand{\cellsecond}{\cellcolor{orange!40}}
\newcommand{\cellthird}{\cellcolor{yellow!40}}

\title{ReCon-GS: Continuum-Preserved Gaussian Streaming for Fast and Compact Reconstruction of Dynamic Scenes}

\author{%
  Jiaye Fu$^{1,2}$, Qiankun Gao$^{1}$, Chengxiang Wen$^{1}$, Yanmin Wu$^{1}$, \\ 
  \textbf{Siwei Ma}$^{2}$, \textbf{Jiaqi Zhang}$^{2\ast}$, \textbf{Jian Zhang}$^{1,3}$\thanks{corresponding to: zhangjian.sz@pku.edu.cn, jqzhang@pku.edu.cn} \\
  $^{1}$School of Electronic and Computer Engineering, Peking University\\
  $^{2}$National Engineering Research Center of Visual Technology, Peking University\\
  $^{3}$Guangdong Provincial Key Laboratory of Ultra High Definition Immersive Media Technology,\\
  Peking University\\
}

\begin{document}

\maketitle

\begin{figure}[htbp]
    \centering
    \begin{subfigure}[b]{0.59\textwidth}
        \centering
        \includegraphics[width=\textwidth]{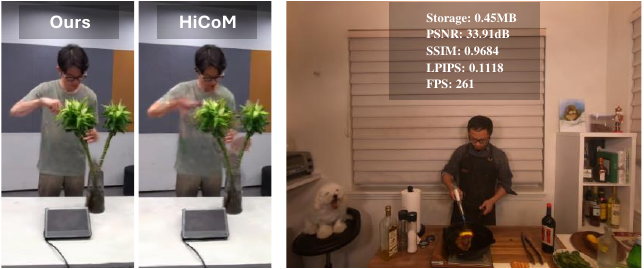}
        \label{fig:teaser:left}
    \end{subfigure}
    \hfill
    \begin{subfigure}[b]{0.37\textwidth}
        \centering
        \includegraphics[width=\textwidth]{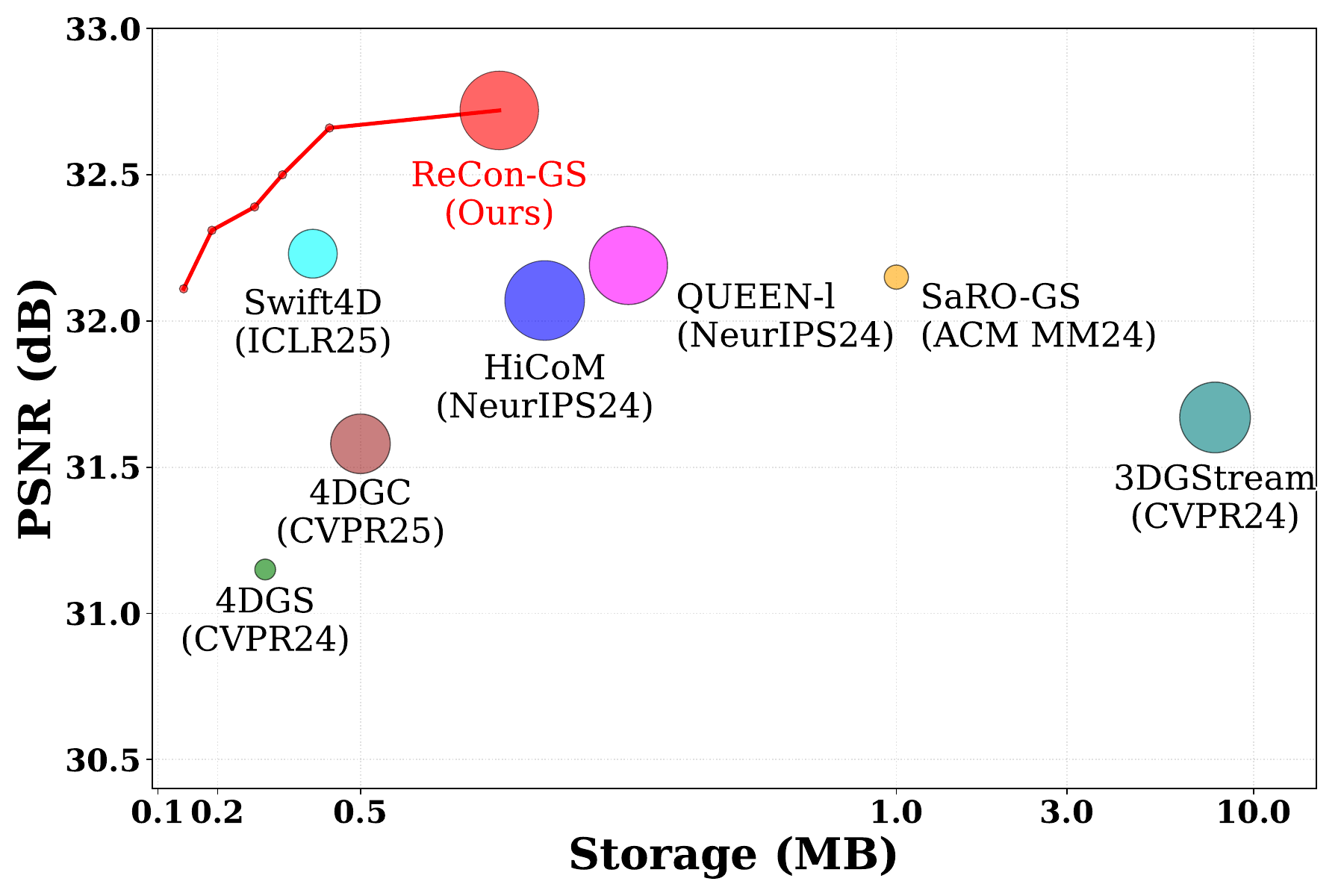}
        \label{fig:teaser:right}
    \end{subfigure}
    
    \vspace{-12pt}
    \caption{The proposed ReCon-GS framework for streamable dynamic scene reconstruction achieves superior rendering qualty with substantially reduced storage. The left figures show the high-quality rendering results of ReCon-GS in a streaming fashion. The right figure is the performance comparison with previous SOTA~\cite{4d-gs,4dgc,hicom,queen,sarogs,3dgstream}, where the radius of circle corresponds to the rendering speed. }
    \label{fig:teaser}
\vspace{-8pt}
\end{figure}

\begin{abstract}
\vspace{-6pt}
%开头
% 自由视点视频（FVV）流式重建面临重大挑战。近期基于流式3D高斯溅射（3DGS）的方法虽展现出前景，但仍存在单帧训练速度慢、运动恢复不稳定及存储成本过高等问题，限制了其在复杂动态场景中的应用。
% Version 1
% Streaming free-viewpoint video (FVV) reconstruction poses significant challenges. Recent streaming 3D Gaussian Splatting (3DGS) methods demonstrate promising progress, yet they still suffer from slow per-frame training, unstable motion recovery, and prohibitive storage costs, thereby limiting their utility in complex dynamic scenarios.
% Version 2
% The reconstruction of streaming free-viewpoint video (FVV) faces several challenges, including slow per-frame optimization, inconsistent motion estimation, and unsustainable storage demands.
% Version 3
Online free-viewpoint video (FVV) reconstruction is challenged by slow per-frame optimization, inconsistent motion estimation, and unsustainable storage demands.
% 提出方法
% 本文提出自适应连续拓扑可重构高斯流（ReACT-GS）——一种面向高保真实时渲染的动态场景重建的新型存储感知框架。 
% Version 1
% This paper introduces the \textbf{Re}configurable \textbf{Con}tinuum Gaussian Stream, termed \textbf{\textbf{ReCon-GS}}, a novel storage-aware framework that enables high-fidelity online dynamic scene reconstruction and real-time rendering.
% Version 2
To address these challenges, we propose the \textbf{Re}configurable \textbf{Con}tinuum Gaussian Stream, dubbed \textbf{ReCon-GS}, a novel storage-aware framework that enables high-fidelity online dynamic scene reconstruction and real-time rendering.
% 方法陈述
% 我们的框架基于高斯元密度自适应的分配多层次锚点高斯来捕捉帧间的几何变形，将运动分解为由粗到细的紧凑表示。同时，为了解决帧间变形导致的锚点高斯运动表达能力下降，我们提出了动态重分配策略（Dynamic Hierarchy Reconfiguration strategy），通过锚点高斯的重层级化（re-hierarchization）保证锚点高斯始终高效的表示局部运动。并且通过利用intra-hierarchical deformation inheritance保证逐层次间的锚点高斯的时域运动一致性。此外，ReACT-GS通过策略性调节各层次锚点高斯密度实现灵活的质量-内存权衡，从而形成存储感知优化机制。 
% Version 1
% Our framework dynamically allocates multi-level anchor Gaussians in a density-adaptive manner to capture inter-frame geometric deformations, decomposing scene motion into coarse-to-fine compact representations.
% To address motion representation degradation caused by the drift of anchor Guassian, we propose a Dynamic Hierarchy Reconfiguration strategy that maintains localized motion expressiveness through on-demand anchor re-hierarchization, while ensuring temporal motion consistency via intra-hierarchical deformation inheritance that propagates transformation priors exclusively within corresponding hierarchy levels. 
% Furthermore, our \textbf{ReCon-GS} framework inherently enables storage-aware optimization by strategically adjusting the density of anchor Gaussians at each level, achieving flexible fidelity-memory trade-offs. 
Specifically, we dynamically allocate multi-level Anchor Gaussians in a density-adaptive fashion to capture inter-frame geometric deformations, thereby decomposing scene motion into compact coarse-to-fine representations.
Then, we design a dynamic hierarchy reconfiguration strategy that preserves localized motion expressiveness through on-demand anchor re-hierarchization,  while ensuring temporal consistency through intra-hierarchical deformation inheritance that confines transformation priors to their respective hierarchy levels.
Furthermore, we introduce a storage-aware optimization mechanism that flexibly adjusts the density of Anchor Gaussians at different hierarchy levels, enabling a controllable trade-off between reconstruction fidelity and memory usage.
% 结果
% 在三个主流数据集上的大量实验表明：相比前沿方法，AUGS提升约15%的训练效率，以更优的鲁棒性和稳定性实现了卓越的自由视点视频合成质量。在同等渲染质量下，AUGS较主流方法降低超过50%的内存需求。
Extensive experiments on three widely used datasets demonstrate that, compared to state‐of‐the‐art methods, \textbf{ReCon-GS} improves training efficiency by approximately 15\% and achieves superior FVV synthesis quality with enhanced robustness and stability. Moreover, at equivalent rendering quality, \textbf{ReCon-GS} slashes memory requirements by over 50\% compared to leading state‑of‑the‑art methods. Code is avaliable at: \url{https://github.com/jyfu-vcl/ReCon-GS/}.
\end{abstract}

\clearpage
\section{Introduction}
\vspace{-5pt}
% 中文对照
% 背景
% 自由视点视频（Free-viewpoint Video, FVV）重建技术已成为新一代沉浸式媒体系统的关键技术，其发展受到扩展现实（XR）、增强现实（AR）与虚拟现实（VR）应用日益增长需求的强力驱动。 尽管该技术能够支持任意视点的动态场景探索，但受限于时空复杂性，FVV在实时性能与重建精度之间的平衡仍面临固有挑战。 传统离线方法虽可实现高质量重建，却无法满足流媒体、沉浸式直播等实时应用的低延迟需求。因此，一些方法采用流式重建框架，通过一帧一帧的建模帧间变形，实现逐帧的重建。

Free-viewpoint Video (FVV) reconstruction has emerged as a cornerstone of next-generation immersive media systems, driven by the escalating demands of XR, AR, and VR applications. 
While this technology enables dynamic scene exploration from arbitrary viewpoints, its inherent spatiotemporal complexity poses persistent challenges in balancing real-time performance with reconstruction accuracy. 
Offline approaches~\cite{4dgs,4d-gs,rotor4dgs,spacetimegaussian}, though capable of high-quality reconstruction, fail to meet the low-latency requirements of real-time applications such as streaming media and immersive live broadcasting. 
To address this limitation, recent methods~\cite{deformabl3dgs,3dgstream} adopt streaming reconstruction frameworks that progressively model inter-frame deformation for frame-by-frame reconstruction.

% 问题描述
% 尽管取得上述进展，实现实用化实时自由视点视频重建仍存在关键瓶颈。
% 首先，当前流式3D高斯溅射方法的变形场带来巨量的冗余存储，这主要是主要是因为高斯基元的均匀空间编码策略与物理运动层级特性本质不兼容，导致支配宏观运动的刚体变换建模失准，引发冗余形变参数存储，同时还导致了严重的误差累计问题。
% 一些方法通过引入额外的辅助方法（如：光流估计网络）以消除流式运动建模带来的误差累计问题，但这会导致训练成本增加以及实用性严重下降的问题。
% 此外，现有框架依赖静态单目标优化范式，僵化地优先保真度或存储效率，缺乏工业动态部署所需的自适应平衡能力——此类场景下计算资源与延迟阈值常随操作环境变化。
Despite these advancements, critical bottlenecks persist in achieving practical real-time FVV reconstruction.
First, current streaming 3D Gaussian splatting methods incur redundant storage due to deformation fields, primarily attributed to the fundamental incompatibility between the uniform spatial motion encoding strategy of Gaussian primitives and the inherent hierarchical characteristics of physical motion. This mismatch leads to inaccurate and inefficient deformation modeling for dominant macroscopic motions, triggering redundant deformation parameter storage and severe error accumulation issues. While some approaches~\cite{motion_aware_gs,igs} mitigate error accumulation in streaming motion modeling by introducing auxiliary mechanisms (\textit{e.g.}, optical flow estimation networks), such solutions inevitably increase training costs and significantly compromise
 practical applicability. 
%Furthermore, existing frameworks rely on static single-objective optimization paradigms that rigidly prioritize fidelity or storage efficiency, lacking the adaptive balancing capability required for industrial dynamic deployment—where computational resources and latency thresholds frequently vary across operational environments.
Furthermore, Industrial deployment necessitates dynamically balancing storage efficiency and fidelity to accommodate fluctuating resources. However, existing frameworks employ static strategies, thereby lacking the flexibility necessary for practical deployment.

% 所题方法
% 针对上述挑战，我们提出\textbf{自适应层次化解耦高斯流}（\textbf{ReCon-GS}），一种联合优化训练效率、渲染保真度与存储自适性的新型流式框架。
% 本方法提出\textbf{自适应层次化运动表征}（AHMR），该锚点驱动的多级运动编码范式通过基于网格的最远点采样将场景动态分解为粗粒度到细粒度的基于锚点的显式运动表达结构，使锚点高斯分布与空间几何特征对齐。同时，通过聚类算法，锚点高斯的显式变形参数共享给聚类中的一般高斯使得AHMR在有效的表征多尺度刚体运动的同时，保证了存储的紧凑性。
% 同时为了解决空间形变导致的锚点高斯偏移从而使得锚点高斯表征局部运动能力下降的问题，ReCon-GS提出重-层级化以及层级变形场，通过锚点高斯轨迹匹配以及层级形变以保持锚点的运动表证能力并保证了锚点变换时的运动连续性。
% 同时，得益于ReCon-GS的框架结构，将4D重建重构为存储感知的保真度优化问题，ReCon-GS实现了类似于率失真权衡理论的范式革新，能够根据应用场景需求动态平衡内存效率与渲染质量，且不损失实时性能。

To address the aforementioned challenges, we propose the \textbf{Re}configurable \textbf{Con}tinuum Gaussian Stream, referred to as \textbf{ReCon-GS}, a novel streaming framework that jointly optimizes rendering fidelity and storage adaptability. 
Our method introduces \textit{Adaptive Hierarchical Motion Representation}, an anchor-driven multi-scale motion encoding paradigm that dynamically allocates Anchor Gaussians through grid-based farthest point sampling to decompose scene dynamics into coarse-to-fine motion structures. 
 This paradigm ensures alignment between Anchor Gaussian distributions and spatial geometric features. 
 Deformation parameters are propagated to clustered General Gaussians via hierarchical mechanisms, enabling effective multi-scale rigid motion representation with minimal storage requirements.
Then, \textbf{ReCon-GS} implements a \textit{Dynamic Hierarchy Reconfiguration} strategy to resolve the degradation of local motion representation caused by Anchor Gaussian drift during spatial deformation.
Specifically, within each hierarchy level, we perform trajectory matching of Anchor Gaussians across frames to track their movement throughout the sequence, while constraining deformation field propagation to operate exclusively within these matched trajectories. 
Reorganized anchors inherit deformation priors only from their original predecessors at the same hierarchy level, thereby preserving both motion inheritance granularity and temporal motion coherence.
Consequently, leveraging these architectural innovations, we reformulate 4D reconstruction as a storage-aware fidelity optimization problem, achieving paradigm-level advancements analogous to rate-distortion optimization theory.  
This enables dynamic balance between memory efficiency and rendering quality according to application demands without compromising real-time performance, as our framework adaptively adjusts the density of Anchor Gaussians at different hierarchy levels based on scene complexity.

Extensive experiments validate our framework’s efficacy across diverse scenarios, demonstrating that \textbf{ReCon-GS} achieves compact yet high-fidelity reconstruction under varying motion patterns while outperforming state-of-the-art streaming methods in three well-known datasets. Our key contributions are summarized as follows:
\begin{itemize}
\item We propose Adaptive Hierarchical Motion Representation, an anchor-driven multi-scale motion encoding paradigm that adaptively align Anchor Gaussians with scene geometry to decompose scene dynamics into coarse-to-fine structures, achieving an efficient yet compact motion representation.

\item We design a Dynamic Hierarchy Reconfiguration strategy to address anchor drift-induced motion degradation, combining re-hierarchization with intra-level deformation inheritance to propagate priors exclusively within original hierarchy layers, eliminating cross-layer interference while preserving motion fidelity.

\item Comprehensive experiments across multiple datasets validate \textbf{ReCon-GS}'s superiority over state-of-the-art streaming methods. Our framework achieves superior performance with 15\% faster training convergence while reducing storage requirements by over 50\% at equivalent rendering quality. Even under constrained storage budgets, \textbf{ReCon-GS} maintains significant rendering quality improvements, demonstrating robust practical efficacy. % FIX-ME
\end{itemize}
\vspace{-4pt}

\section{Related Work}
\label{sec::related_work}
\subsection{3D Gaussian Splatting for Static Scenes}
\vspace{-5pt}
% 随着新视角合成任务的兴起，3D高斯泼溅得益于其高渲染质量以及快速地训练逐渐取代基于神经辐射场的相关方法成为主流的静态场景重建任务的方法。然而，巨量的存储使得极大地影响了3DGS的实用性。因此，很多的研究都聚焦于如何在保证高质量渲染的同时实现更紧凑的3DGS表示。一些方法专注于裁剪高斯元的数量，例如通过可训练掩码机制去除不重要的高斯元，通过梯度阈值的方式等。此外，一些方法专注于减少3D高斯的参数从而减小每个3DGS的存储。而，基于Vector Quatization理念的3DGS方法也被证明可以有效的减小3DGS的存储。其他的方法则是专注于探索紧凑的空域结构表示以减少3DGS的存储。Scaffold-GS通过聚类高斯元到锚点并使用MLP进行每个聚类的高斯元属性隐式表示有效的降低了高斯元的存储。后续的工作通过将空域结构先验注入熵编码器等压缩手段从而进一步压缩3dGS。

With the rise of novel view synthesis (NVS) tasks, 3D Gaussian Splatting (3DGS)~\cite{3dgs} has gradually replaced Neural Radiance Fields (NeRF)-based methods~\cite{nerf,nerf++,mipnerf,zipnerf,mipnerf360} as the mainstream approach for static scene reconstruction due to its high rendering quality and fast training speed. However, its massive storage requirements significantly hinder the practical application of 3DGS. Consequently, numerous studies have focused on achieving more compact 3DGS representations while maintaining high-quality rendering. Some methods~\cite{lightgaussian,lpgs,eagles} concentrate on pruning Gaussian primitives, such as removing unimportant Gaussians through trainable mask mechanisms~\cite{compact3dgs} or gradient thresholding approaches~\cite{elmgs}. Additionally, other approaches~\cite{compact3d} aim to reduce the parameters of 3D Gaussians to decrease per-primitive storage. Meanwhile, 3DGS methods~\cite{lightgaussian,compact3dgs,compact3d,compressedgs,e2egs} based on Vector Quantization (VQ) principles have also proven effective in reducing storage demands. Alternative strategies~\cite{compgs,lags,cat3dgs} explore compact spatial structure representations to minimize 3DGS storage. Scaffold-GS~\cite{scaffoldgs} achieves efficient storage reduction by clustering Gaussians to anchor points and implicitly representing Gaussian attributes for each cluster through MLPs. Subsequent works~\cite{hac,contextgs,hac++} further compress 3DGS by incorporating spatial structure priors into entropy encoders and other compression techniques.
\vspace{-5pt}
\subsection{3D Gaussian Splatting of Dynamic Scenes}
\vspace{-5pt}
% 3D动态重建任务是计算机视觉、图形学领域中更为挑战性的任务。随着3DGS的兴起，3DGS的动态重建逐渐演变出两种路径，一种是通过构建高斯标准场以及时间与变形场的隐式表示实现高效的表示的离线方法。然而，离线方法如何进行高效的运动表示始终是一个问题，一些方法通过锚点、控制点的方式通过将高斯元嵌入到一种更为稀疏的表示从而实现更为高效的运动表示。
% 另一种动态重建的范式则是使用3DGS进行首帧的高质量后，通过构建变形场逐帧通过变形后的3DGS进行表示，从而实现了on-the-fly的动态重建框架。然而，每帧的运动场表示带来的巨量存储开销让流式框架面临极大的挑战。Gao等人通过提出分层的运动表达框架，实现了由粗到细的3d高斯变形表示，有效的降低了变形场的开销。同时，流式框架由于其框架特性面临着严重的误差累计问题，一些方法通过使用预训练的运动估计模型通过逐帧的二维图像的运动指导高效的3D高斯变形表示从而提升整体渲染质量。
% 通过向4DGS中引入锚点、控制点的机制从而对高斯元属性进行紧凑表征，【4D-Scaffold-GS、MoDec-GS】有效的对变形场表征进行压缩。【Saro-GS】通过引入缩放感知的残差场实现了变形场关于高斯元时-空关系的紧凑建模，从而实现了高帧率的渲染效果。【4DGC】通过将前一帧的高斯元信息作为先验引入熵模型从而实现了紧凑的高斯元压缩。另一种动态重建的范式则是使用3DGS进行首帧的高质量后，通过构建变形场逐帧通过变形后的3DGS进行表示，从而实现了on-the-fly的动态重建框架【3dgstream】。【HiCoM】通过提出分层的运动表达框架，实现了由粗到细的3d高斯变形表示。【IGS】通过使用预训练的运动估计模型通过逐帧的二维图像的运动指导高效的3D高斯变形表示从而提升整体渲染质量。
3D dynamic scene reconstruction stands as one of the most challenging tasks in computer vision and graphics. With the rise of 3D Gaussian Splatting (3DGS), dynamic reconstruction approaches have diverged into two main paradigms. The first involves offline methods~\cite{4dgs,4d-gs,rotor4dgs,fedgs,dn4dgs} that implicitly model scene dynamics through canonical Gaussian fields coupled with spatiotemporal deformation fields. While these methods achieve compact representations, they struggle with efficient motion parameterization. Recent advancements address this by embedding Gaussians into sparse motion representations using anchor points~\cite{4dscaffold,mosca,modecgs} or control points~\cite{sc-gs,sk-gs,sp-gs}, enabling motion propagation through sparse keyframes while preserving reconstruction quality.
The second paradigm adopts streaming frameworks that first reconstruct a high-quality 3DGS representation for the initial frame and then model subsequent frames via deformation fields~\cite{dynamic3dgs,3dgstream,deformabl3dgs}. However, the massive storage overhead of per-frame motion parameters poses significant challenges. Gao et al.~\cite{hicom} mitigate this through a hierarchical motion representation framework, decomposing deformations into coarse-to-fine transformations to reduce parameter dimensionality. Despite this, streaming frameworks inherently suffer from error accumulation due to incremental updates. To address this, some methods~\cite{motiongs,gaufre,motion_aware_gs,igs,compact-d3dgs} leverage pre-trained motion estimation models (e.g., optical flow networks) to guide 3D Gaussian deformations via 2D motion priors, enhancing both rendering quality and temporal consistency. These innovations highlight the critical balance between storage efficiency, computational tractability, and reconstruction fidelity in dynamic 3D modeling.
\vspace{-4pt}

\begin{figure}[t]
  \centering
  \includegraphics[width=\textwidth]{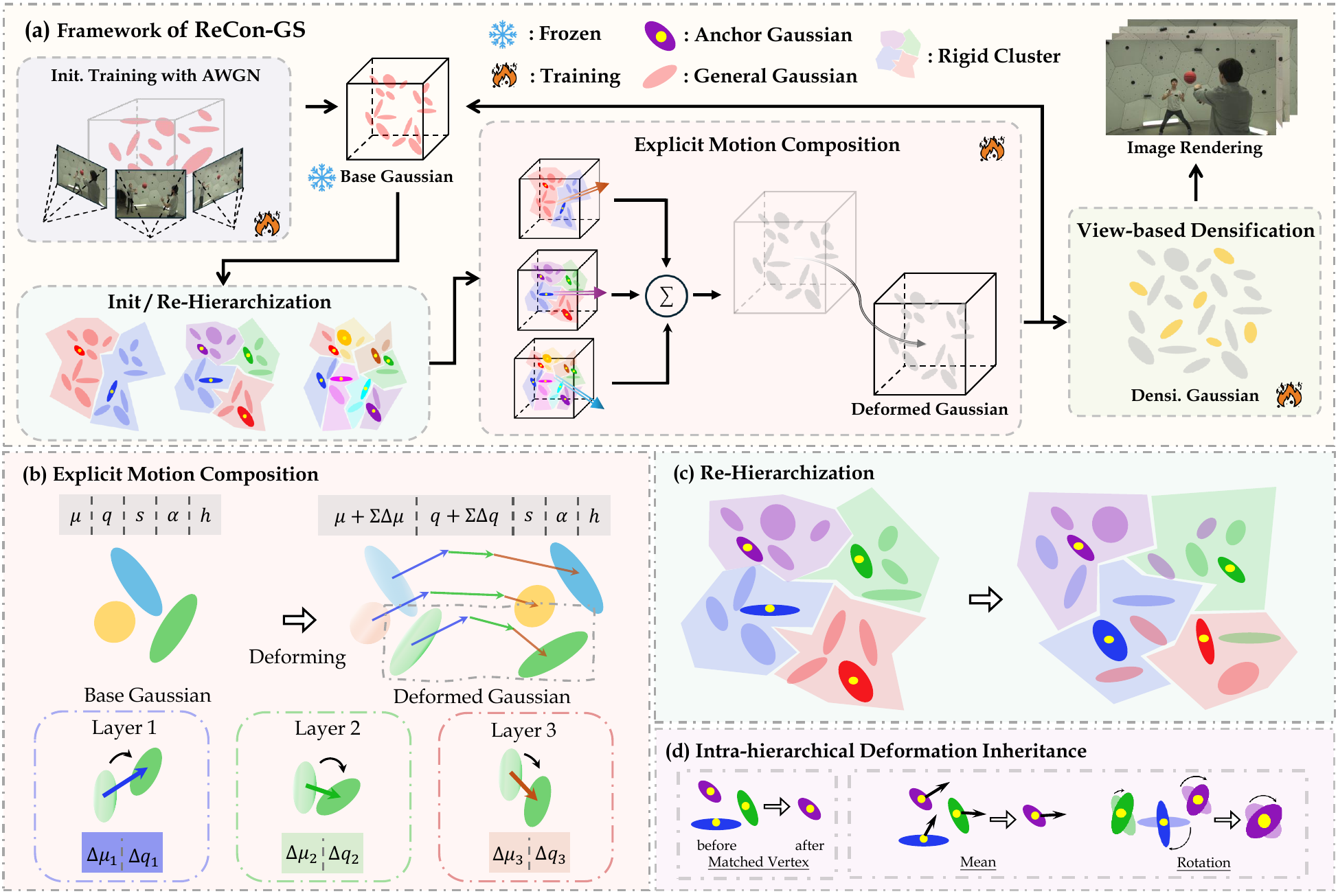}
  \caption{
    \textbf{Illustration of our ReCon-GS framework.}
    (a) We begin by generating a high‐quality base 3D Gaussian Splatting (3DGS) representation for the first frame, then embed it into an adaptively hierarchical motion presentation framework via grid‐based farthest‐point sampling.
    (b) Explicit Motion Composition updates the base 3DGS across successive frames.
    (c) and (d) Periodically, a Re‐Hierarchization stage accommodates complex object motion while preserving temporal consistency through Intra-hierarchical Deformation Inheritance. 
    Finally, a view‐based densification further refines the 3DGS for high‐quality rendering.
  } 
  \label{fig:framework}
\end{figure}

\section{Preliminaries}
\vspace{-5pt}
% More concise version:
3DGS models scenes using anisotropic 3D Gaussians defined by a mean $\mathbf{\mu}$ and a covariance matrix $\mathbf{\Sigma}=\mathbf{R} \mathbf{S} \mathbf{S}^\top \mathbf{R}^\top$ , where $\mathbf{S}=diag(s_x,s_y,s_z)\in \mathbb{R}^3$ represents axis-aligned scaling, and $\mathbf{R}\in SO\left(3 \right)$ is parameterized by a quaternion $\mathbf{q}$. Each Gaussian's view-dependent color uses spherical harmonics $\mathbf{SH}$ and opacity $\mathbf{\alpha}\in \left[ 0,1 \right]$.

The mathematical formulation of Gaussian is:
\begin{equation}
\mathcal{G}(\mathbf{x}) = e^{-\frac{1}{2}(\mathbf{x} - \boldsymbol{\mu})^\top \mathbf{\Sigma}^{-1}(\mathbf{x} - \boldsymbol{\mu})}.
\label{sec:pre:eq:gaussian_definition}
\end{equation}

% 3DGS~\cite{3dgs} utilizes anisotropic 3D Gaussians to explicitly model scenes. Each Gaussian $\mathbf{\mathcal{G}}$ is defined by a mean vector $\mathbf{\mu}$  and a covariance matrix $\mathbf{\Sigma}$, which jointly specify the spatial centroid and anisotropic geometry of the Gaussian in world coordinates. The mathematical formulation is:
% \begin{equation}
% \mathcal{G}(\mathbf{x}) = e^{-\frac{1}{2}(\mathbf{x} - \boldsymbol{\mu})^\top \mathbf{\Sigma}^{-1}(\mathbf{x} - \boldsymbol{\mu})}.
% \label{sec:pre:eq:gaussian_definition}
% \end{equation}

% To ensure physical validity and differentiable optimization, the covariance matrix $\mathbf{\Sigma}$ is decomposed into a scaling matrix $\mathbf{S}$ and a rotation matrix $\mathbf{R}$ via: 
% \begin{equation}
% \mathbf{\Sigma} = \mathbf{R} \mathbf{S} \mathbf{S}^\top \mathbf{R}^\top,
% \label{sec:pre:covariance_decomposition}
% \end{equation}
% where $\mathbf{S}=diag(s_x,s_y,s_z)\in \mathbb{R}^3$ represents axis-aligned scaling, and $\mathbf{R}\in \mathbf{SO}\left(3 \right)$ is parameterized by a quaternion $\mathbf{q}$. Each Gaussian is further augmented with view-dependent color properties encoded by spherical harmonics coefficients $\mathbf{SH}$ and an opacity parameter $\mathbf{\alpha}\in \left[ 0,1 \right]$.

For novel view synthesis, Gaussians are projected onto the imaging plane using a view transformation matrix $\mathbf{W}$. The projected 2D covariance $\mathbf{\Sigma}'$ is computed through the Jacobian $\mathbf{J}$ of the projective transformation:
\begin{equation}
\mathbf{\Sigma}' = \mathbf{J} \mathbf{W} \mathbf{\Sigma} \mathbf{W}^\top \mathbf{J}^\top.
\label{eq:projected_covariance}
\end{equation}
Rendering proceeds by alpha-blending $N$ depth-ordered Gaussians at each pixel, with the final color calculated as:
\begin{equation}
C = \sum_{i \in N} \mathbf{c}_i \alpha_i \prod_{j=1}^{i-1} (1 - \alpha_j),
\label{eq:pixel_color}
\end{equation}
where $\mathbf{c}_i$ and $\alpha_i$ correspond to the color and blending weight of the $i^{th}$ Gaussian, respectively.

The learning pipeline iteratively alternates between gradient-based parameter updates and topological adaptation. Parameter optimization is supervised by the $\mathcal{L}_1$ loss and D-SSIM term:
\begin{equation}
    \mathcal{L}=(1-\lambda)\mathcal{L}_1+\lambda\mathcal{L}_\text{D-SSIM},
\end{equation}
where $\lambda$ is typically set to 0.2.
\vspace{-4pt}

\section{Methodology}
\subsection{Overview}
\vspace{-5pt}
% 我们的目标是通过多视角视频在不同的存储需求下实现高质量的动态场景。为了实现这个目标，我们采用了自适应的分层运动表征策略实现运动，通过将高斯元的运动分解成三个不同尺度的刚体运动的集合，每个刚体的位置由每层通过基于网格的最远点采样得到的锚点高斯的位置决定。刚体内部的其余高斯元均共享锚点高斯的位置。为了进一步让我们基于锚点的运动表达适配不同运动幅度的场景，我们通过引入重-层级化的策略通过重新分配锚点高斯以适应随着时间不短变化的场景几何。完整的框架在图~\ref{}中展示。

% 我们的模型首先跟随HiCoM使用了基于加性噪声注入的初始帧3DGS训练方法生成了紧凑的基础高斯元。随后，为了高效地表示高斯元在每帧的变形，我们采用了自适应的分层运动表征策略实现运动，通过将高斯元的运动分解成三个不同尺度的刚体运动的集合，每个刚体的位置由每层通过基于网格的最远点采样得到的锚点高斯的位置决定。刚体内部的其余高斯元均共享锚点高斯的位置。为了进一步让我们基于锚点的运动表达适配不同运动幅度的场景，我们通过引入重-层级化的策略通过重新分配锚点高斯以适应随着时间不短变化的场景几何。最后，为了进一步refine局部细节，通过基于视角的密集化过程进一步调整3DGS以渲染更高质量的场景。完整的框架在图~\ref{}中展示。
Our goal is to achieve high-quality dynamic scene reconstruction from multi-view videos under varying storage constraints. To accomplish this, our framework first generates compact Base Gaussians $\mathcal{G}_B$ through an additive noise-injected 3DGS training strategy~\cite{hicom} during initial frame reconstruction. Subsequently, to efficiently parameterize per-frame Gaussian deformations, we propose an adaptive hierarchical motion representation framework that explicitly models scene dynamics through scale-aware motion decomposition. Specifically, each Gaussian's motion is parameterized as a composition of layer-wise rigid transformations across multiple geometric scales, with each deformation layer governed by a distinct set of Anchor Gaussians~$\mathcal{G}_A$ spatially distributed via grid-based farthest-point sampling. General Gaussians within each layer hierarchically inherit blended motion parameters from their nearest anchor counterparts through cross-scale deformation coefficients, enabling progressive motion aggregation from coarse to fine hierarchies.

To enhance the adaptability of our anchor-based motion representation for scenes with varying motion magnitudes, we introduce dynamic hierarchy reconfiguration - a self-optimizing mechanism that redistributes Anchor Gaussians to match evolving scene geometry while maintaining temporal motion consistency through Intra-hierarchical Deformation Inheritance. Finally, to further refine local details, ReCon-GS optimizes 3DGS representation through a view-adaptive densification process, enhancing scene rendering fidelity. The complete pipeline is illustrated in Figure~\ref{fig:framework}~(a).

\subsection{Adaptively Hierarchical Motion Presentation}
\label{sec:metd:ahmp}
Inspired by the hierarchical decomposition of real-world object motion into multi-scale rigid-body components, we propose an adaptive hierarchical motion representation framework. 
Our framework begins with the \textbf{Base Gaussians} $\boldsymbol{\mathcal{G}_{B}}$, which is the static 3D Gaussians reconstructed from the first frame of multi-view inputs. Building upon $\mathcal{G}_{B}$, we establish an adaptive hierarchical motion representation through three core stages:

\textbf{Anchor Gaussian Initialization}. Given $\mathcal{G}_B=\{ g_i \}^{N}_{i=1}$ with position $\mu_{i}\in\mathbb{R}^3$, we first compute the scene's spatial bounds as $\mu_{min}=\min_{i}\mu_{i}$ and $\mu_{max}=\max_{i}\mu_{i}$. $\mathcal{G}_B$ is then divided into a uniform 3D grid with resolution $M=\left\lceil N_{anchor}^{1 / 3}\right\rceil$, where $N_{anchor}$ is the number of anchor. For each cell $\mathcal{C}_{i,j,k}$, we select the \textbf{Anchor Gaussians} $g_a \in \boldsymbol{\mathcal{G}_{A}}$ with position $\mu_a$ closest to its geometric center:
\begin{equation}
\mathbf{g}_{a}=\arg \min _{\mathbf{g} \in \mathcal{C}_{i, j, k}}\left\|\boldsymbol{\mu}-\mathbf{c}_{i, j, k}\right\|_{2}, \ \ where \ \ \mathbf{c}_{i, j, k}=\boldsymbol{\mu}_{\min }+\left(i+\frac{1}{2}, j+\frac{1}{2}, k+\frac{1}{2}\right) \odot \frac{\boldsymbol{\Delta}}{M},
\end{equation}
with  $\Delta=\mu_{max}-\mu_{min}$. This \textit{grid-based farthest point sampling} ensures each anchor locally represents its cell while maintaining global uniformity. Compared to conventional farthest point sampling approaches, our grid-based strategy achieves significant computational acceleration through structured sampling constraints, effectively reducing the time complexity while preserving sampling quality.

\textbf{Hierarchical Rigid-Cluster Formation}. With $\mathcal{G}_A$ initialized, each \textbf{General Gaussians} $g_n\in\mathcal{G}_N$ is assigned to its nearest anchor via L1-distance minimization:
\begin{equation}
g_a^*=\arg\min_{g_a\in\mathcal{G}_A} \| \mu_n-\mu_a \|_1.
\end{equation}
forming primary rigid clusters $\{ \mathcal{R}_a \}$.  To address residual motions within rigid clusters, we recursively subdivide the hierarchy into two finer levels within the original spatial coordinate system. At each subsequent level, we increase the density of Anchor Gaussians by adapting the grid resolution $M=\left\lceil (N_{anchor} \, \cdot \, 3^{l-1})^{1 / 3}\right\rceil$, where $l={2,3}$ denotes the hierarchy level. 

\textbf{Explicit Motion Composition}. As shown in Figure~\ref{fig:framework}~(b), each hierarchy level $l\in\{ 1, 2,3 \}$ associates clusters with rigid transformations $\mathbf{T}^{(l)}=(\Delta \mathbf{\mu} ^ {(l)},\Delta \mathbf{q} ^ {(l)},)$, where $\Delta \mathbf{\mu} ^ {(l)} \in \mathbb{R}^3$ and $\Delta \mathbf{q} ^ {(l)} \in \mathbb{R}^4$ denote incremental translation and rotation of Anchor Gaussians $\mathcal{G}_A$. Thus, Anchor Gaussians $\mathcal{G}_A$ act as rigid motion keypoints that govern the affine transformation of all General Gaussians $\mathcal{G}_N$ within their associated cluster.
 By propagating these parameters hierarchically, the motion of each Anchor dictates the collective deformation of its rigid sub-region, emulating real-world rigid-body dynamics.
The total deformation of a Gaussian is the cumulative sum across levels: 
\begin{equation}
\Delta \boldsymbol{\mu}_{h} = \sum_{l=1}^3 \Delta \boldsymbol{\mu}^{(l)},\quad \Delta \boldsymbol{q}_{h} = \sum_{l=1}^3 \Delta \boldsymbol{q}^{(l)}.
\end{equation}
Upper level model global rigid motions shared across clusters, while the finest level resolves residual deformations unique to individual Gaussians. Crucially, we decouple geometric attributes (optimized via $\Delta \boldsymbol{\mu}_{h}$ and $ \Delta \boldsymbol{q}_{h}$) from static appearance parameters (sphere harmonic coefficient $SH$ and opacity $\alpha$ ), freezing the latter to avoid parameter explosion from separately modeling appearance deformations, which is critical for ReCon-GS's storage efficiency. Since pixel intensities rely on $\alpha$-blended Gaussian contributions, precise geometric deformation alone ensures photorealistic rendering. By optimizing geometry while freezing appearance attributes, ReCon-GS achieves parameter efficiency without sacrificing visual quality.

\textbf{Motion Parameter Preservation.} ReCon-GS can store motion parameters at each level more compactly compared to previous streaming reconstruction method~\cite{hicom}. This is primarily due to the use of a carefully designed \textit{grid-based farthest point sampling} algorithm for anchor Gaussian assignment, instead of a uniform anchor sampling strategy. As a result, ReCon-GS can store motion parameters for Anchor Gaussians based solely on their relative positions, without the need for additional indices. This significantly reduces the memory required for storing motion parameters.
% 对于每层运动参数的保存，ReCon-GS可以比之前流式重建方法更加紧凑。这主要得益于锚点高斯的分配是使用精心设计的基于网格的最远点采样算法得到的，因此ReCon-GS对于锚点高斯的运动参数的保存与提取可以完全依照相对位置而无需保存额外的索引。这减小了运动参数保存的内存。

\subsection{Dynamic Hierarchy Reconfiguration}
\label{sec:metd:dhr}

Using Adaptive Hierarchical Motion Representation framework, we efficiently allocates explicit motion parameters according to scene density and geometric structures. However, ReCon-GS also inherits a critical challenge common to streaming reconstruction methods~\cite{3dgstream, hicom}: persistent scene motion induces severe deformations in local rigid clusters, causing Anchor Gaussians to gradually lose their representativeness as rigid motion bases. This degradation propagates errors through the motion hierarchy, accumulating artifacts in rendered sequences. To address this, we propose Dynamic Hierarchy Reconfiguration strategy—a self-correcting mechanism that periodically reinitializes the hierarchical structure while preserving temporal motion coherence through intra-hierarchical deformation inheritance.

\textbf{Periodic Re-Hierarchization}. At each reconfiguration step $t=kT \ (k\in\mathbb{N}^+)$, the framework regenerate Anchor Gaussians $\mathcal{G}_A$ using grid-based farthest point sampling on the current Gaussian distribution, ensuring updated anchors align with the latest scene geometry. Then, General Gaussian $\mathcal{G}_N$ is reassigned to new anchors via L1-distance minimization. The process is described in Figure~\ref{fig:framework}~(c).

\textbf{Intra-hierarchical Deformation Inheritance}.As present in Figure~\ref{fig:framework}~(d), after re-hierarchization, we implement intra-hierarchical deformation inheritance to enforce temporal motion consistency between $\mathcal{G}_A$ across hierarchy levels. For each hierarchy level $l\in\{  1,2,3 \}$, each new Anchor Gaussian inherits deformation parameters from its three nearest legacy Anchors to ensure temporal continuity. Specifically, The translational deformation is governed by:
\begin{equation}
\Delta{\boldsymbol{\mu}_{a}'}^{(l)} = \frac{1}{3} \sum_{i=1}^3 \Delta\boldsymbol{\mu}_{a_{i}}^{(l)},
\label{sec:metd:eq:trans_inter}
\end{equation}
where $\Delta{\boldsymbol{\mu}_{a}'}^{(l)}$ denotes the translational deformation parameter of new Anchor Gaussian at level $l$, and $\Delta\boldsymbol{\mu}_{a_{i}}^{(l)}$ represents the translational deformation parameter of the $i$-th matched Anchor Guassian.
Rotational deformation follows an eigenvector-based formulation:
\begin{equation}
\Delta{\boldsymbol{q}_{a}'}^{(l)} = \frac{v_{\max}(\mathcal{M})}{\bigl\| v_{\max}(\mathcal{M}) \bigr\|},\quad\mathcal{M} = \sum_{i=1}^{3} \Delta\boldsymbol{q}_{a_{i}}^{(l)}  \bigl(\Delta\boldsymbol{q}_{a_{i}}^{(l)}\bigr)^\mathsf{T},
\label{sec:metd:eq:rot_inter}
\end{equation}
where $\Delta{\boldsymbol{q}_{a}'}^{(l)}$ is the rotational deformation parameter of new Anchor Guassian, and $\Delta\boldsymbol{q}_{a_{i}}^{(l)}$ represents the rotational deformation parameter of the $i$-th matched Anchor Guassian. $v_{max}\left( M \right)$ denotes the eigenvector corresponding to the largest eigenvalue of $\mathcal{M}$.

\subsection{Storage-aware Optimization}
\label{sec:metd:sfo}
% 得益于流式框架的特性，高斯元的存储开销往往只占总存储中很小的一个部分，大部分的存储开销都来自于每帧的变形场参数存储。由于ReCon-GS采用基于锚点的自适应层级运动框架，每层的锚点数量调节使得ReCon-GS成为了存储感知的框架。ReCon-GS可以根据锚点数量调节自适应的适配不同场景下的存储需求，由此，我们将通常的4D重建任务转换成基于不同存储下的渲染质量最优化问题。即：
% （无需翻译，一堆描述渲染质量最优化的公式）
% 。。

% 此外，需要注意的是，ReCon-GS采纳两阶段的训练策略，在第一个训练阶段，我们单独对变形场进行训练。一旦我们获得了已经足够好的变形场，此时基础高斯元已经基本适配了新的运动场景几何，然后我们在第二阶段采用HiCoM的策略进一步优化局部细节以消除 persistent discrepancies between learned motion and scene dynamics
Thanks to the streaming architecture of ReCon-GS, the storage overhead of Gaussian primitives constitutes only a minor fraction of the total storage consumption, with the majority allocated to per-frame deformation field parameters. By adopting an anchor-driven adaptive hierarchical motion representation, ReCon-GS operates as a storage-aware framework—the hierarchical anchor resolution $\{ N^{(l)}_{anchor}\}^3_{l=1}$ dynamically adapts to varying scene complexity and storage constraints. Moreover, ReCon-GS adopts a decoupled two-stage optimization pipeline to balance motion fidelity and computational efficiency. In Phase 1, we exclusively train the hierarchical deformation fields while freezing the geometric and appearance attributes of the Base Gaussians $\mathcal{G}_B$. Once the deformation fields converge, Phase 2 activates the view-based densification to address persistent discrepancies between learned motion and scene dynamics.

\section{Experiments}
\vspace{-5pt}
% 在这个部分，我们在第5.1和5.2章提供了我们的部署以及数据集的细节.对于我们实验结果的完整分析被展示在第5.2张，而第5.4章进行了我们方法的消融实验。结果显示我们的方法在存储以及渲染质量上实现了sota的表现。
% This section presents the dataset specifications and implementation details in Sec.~\ref{sec:experiment:dataset} and Sec.~\ref{sec:experiment:imple}. A comprehensive analysis of our experimental results is provided in Sec.~\ref{sec:experiment:results}, while Sec.~\ref{sec:experiment:ablation} systematically validates our method through ablation studies. Quantitative and qualitative evaluations demonstrate that our approach achieves state-of-the-art performance in both storage efficiency and rendering quality, outperforming existing baselines by significant margins.
\begin{table}[!t]
\footnotesize
\caption{
\textbf{Quantitative comparison} on N3DV dataset. The storage metric includes the size without and with the initial frame, separated by ``/''.  The training time metric includes the first frame training. The method with $^\dag$ is reproduced by us through official code in the same experimental environment.
}
\label{tab:benchmark1}
\centering
\setlength{\tabcolsep}{2.5pt}
\begin{tabular}{l|l|ccccccc}
\toprule
Category & Method & PSNR (dB)$\uparrow$ & SSIM$\uparrow$ & LPIPS$\downarrow$ & Storage (MB)$\downarrow$ & Train (sec)$\downarrow$ & Render (FPS)$\uparrow$\\
      \midrule
\multirow{4}{*}{Offline}
& 4DGS$^\dag$~\cite{4d-gs}     & 31.36 & \cellthird 0.950 & \cellthird 0.131  & \cellbest 0.3 & 7.8 & 30 \\
& STG~\cite{spacetimegaussian} & 32.05 & 0.948 & - & 0.67 & 20 & 140 \\
& SaRO-GS~\cite{sarogs}        & 32.15 & - & -  & 1.0 & - & 40 \\
& Swift4D~\cite{swift4d}       & \cellthird 32.23 & - & - & \cellsecond 0.4 & \cellbest 5.0 & 125 \\
& SplineGS~\cite{splinegs}     & \cellsecond 32.60 & - & - & - & 11 & 76 \\
\midrule
\multirow{7}{*}{Online}
& Dynamic 3DGS~\cite{dynamic3dgs} & 30.67 & - & - & -/9.2 & 560 & - \\
& StreamRF~\cite{streamrf}     & 30.68 & 0.930 & - & 17.7/31.5 & 15 & 12 \\
& 3DGStream$^\dag$~\cite{3dgstream} & 31.35 & 0.948 & \cellsecond 0.130 & 7.6/7.8 & 8.1 & 245 \\
& 4DGC~\cite{4dgc}             & 31.58 & 0.943 & -  & -/0.5 & 50 & 168 \\
& QUEEN-l~\cite{queen}         & 32.19 & 0.946 & 0.136 & -/0.75 & 7.9 & \cellthird 248 \\
& HiCoM$^\dag$~\cite{hicom}    & 32.08 & \cellsecond 0.953 & \cellsecond 0.130 & 0.48/0.69 & \cellthird 6.6 & \cellbest 255 \\
& \textbf{ReCon-GS} (ours)                         & \cellbest 32.66   & \cellbest 0.957 & \cellbest 0.123 & \cellthird 0.40/0.44 & \cellsecond 6.4 & \cellsecond 250 \\
      \bottomrule
    \end{tabular}
\end{table}
\begin{table}[!t]
\footnotesize
\caption{
\textbf{Quantitative comparison} on Meet Room and PanopticSports datasets. The storage metric includes the size without and with the initial frame, separated by ``/''. The training time metric includes the first frame training. The method with $^\dag$ is reproduced by us through official code in the same experimental environment. The method marked with $^{\ast}$ adopts the \textit{Discussion} scene from the Meet Room dataset as the training set, with the remaining scenes serving as the test set. }
\label{tab:benchmark2}
\centering
\setlength{\tabcolsep}{3.5pt}
\begin{tabular}{l|cccccccccc}
\toprule
\multirow{3.5}{*}{Method} & \multicolumn{5}{c}{Meet Room} & \multicolumn{5}{c}{PanopticSports} \\
\cmidrule(lr){2-6} \cmidrule(lr){7-11}
& PSNR & LPIPS & Storage & Train & Render & PSNR &  LPIPS & Storage & Train & Render \\
& (dB)$\uparrow$ & $\downarrow$ & (MB)$\downarrow$  & (sec)$\downarrow$  & (FPS)$\uparrow$ & (dB)$\uparrow$  & $\downarrow$ &  (MB)$\downarrow$  & (sec)$\downarrow$  & (FPS)$\uparrow$     \\
\midrule
3DGStream$^\dag$~\cite{3dgstream}           & 29.30 & 0.188 & 4.0/4.1 & 4.77 & \cellbest 260 & 23.02 & 0.187 & 7.9/8.1 & \cellbest 5.87 & \cellsecond 369 \\          
IGS-l$^{\ast}$~\cite{igs}       & \cellsecond 30.13 & - & -/1.26 & \cellbest 2.67 & 252 & - & - & - & - & - \\
HiCoM$^\dag$~\cite{hicom}  & 29.57 & \cellsecond 0.182 & \cellsecond 0.30/0.39 & 3.91 & 236 & \cellsecond 29.17 & \cellsecond 0.142 & \cellsecond 1.33/2.11 & 8.60 & 358 \\
\textbf{ReCon-GS} (ours)                        & \cellbest 30.84 & \cellbest 0.163 & \cellbest 0.28/0.30 & \cellsecond 3.86 & \cellsecond 256 & \cellbest 29.33 & \cellbest 0.136 & \cellbest 0.64/0.8 & \cellsecond 7.14 & \cellbest 410 \\
\bottomrule
\end{tabular}
\end{table}
% 十字星的是我们用作者开源的代码重新跑的，作者代码中对原始数据集进行了处理，我们和以往的方法一样采用原始数据集，初始点云的生成参考的是4dgs的
% 关于时间问题
% 1. 影响时间的因素包括：代码、高斯数量、实验环境等
% 2. 我们报告的是多次实验的均值
% 时间考虑用4090的结果，相对比较稳定
\begin{table}[ht]
\footnotesize
\caption{
\textbf{Quantitative comparison} on Technicolor dataset. The storage metric includes the size with the initial frame.  The training time metric includes the first frame training. 
}
\label{tab:technicolor_benchmark}
\centering
\setlength{\tabcolsep}{8pt}
\begin{tabular}{l|l|ccccc}
\toprule
Category & Method & PSNR (dB)$\uparrow$ & SSIM$\uparrow$ & Storage (MB)$\downarrow$ & Render (FPS)$\uparrow$\\
\midrule
NeRF-based
& HyperReel~\cite{hyperreel}   & 31.80 & 0.906 & 1.20 & 4 \\ 
\midrule
\multirow{2}{*}{Offline}
& STG~\cite{spacetimegaussian} & \cellsecond 33.60 & -     & \cellsecond 1.10 & \cellsecond 87 \\
& Ex4DGS~\cite{fedgs}          & 33.62 & \cellsecond 0.916 & 2.81 & 72 \\
\midrule
\multirow{2}{*}{Online}
& E-D3DGS~\cite{pergs}         & 33.24 & 0.907 & 1.54 & 79 \\
& \textbf{ReCon-GS} (ours)     & \cellbest 33.83 & \cellbest 0.932 & \cellbest 0.82 & \cellbest 207 \\
\bottomrule
\end{tabular}
\end{table}
\begin{figure}[t]
    \centering
    \begin{subfigure}[b]{0.48\textwidth}
        \centering
        \includegraphics[width=\textwidth]{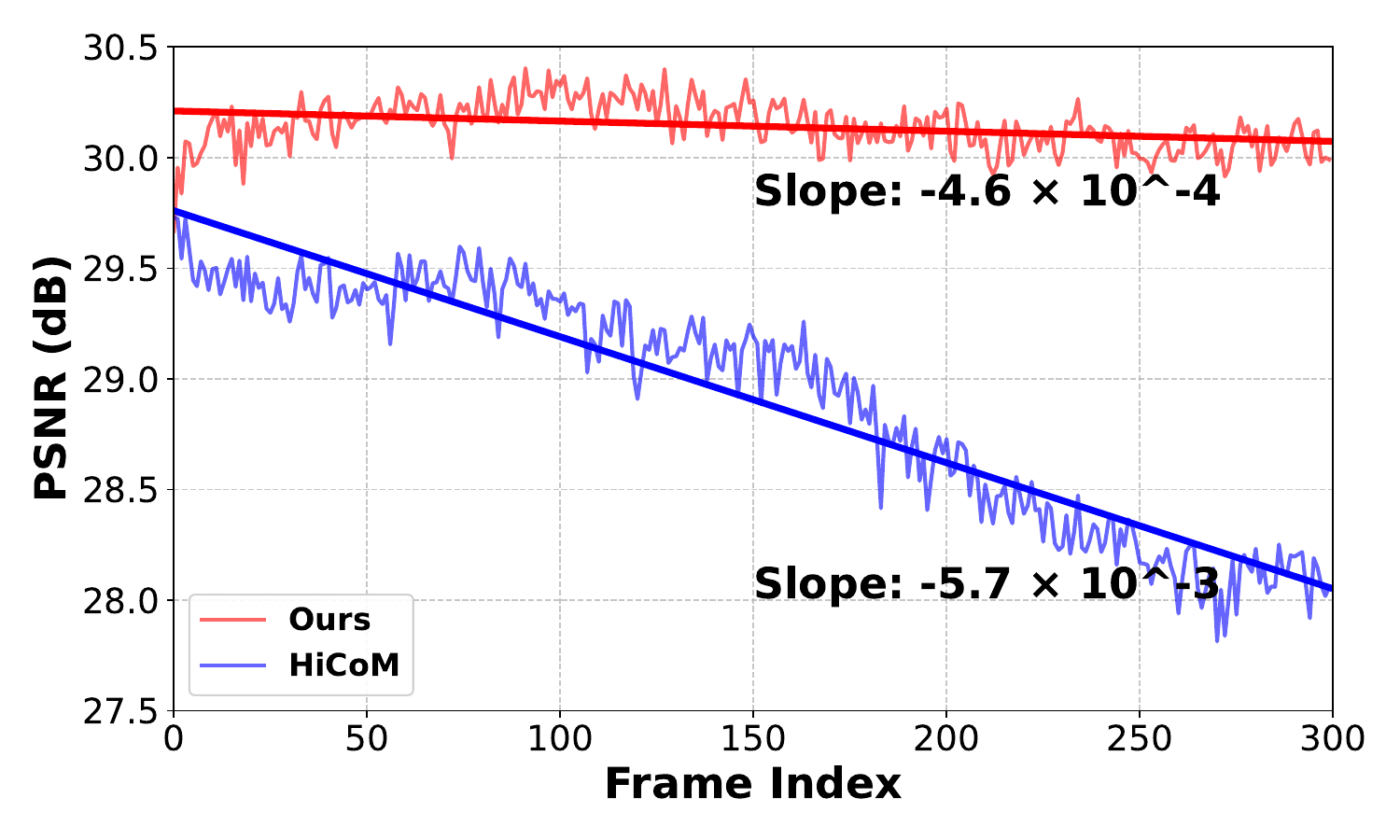}
        \caption{\textit{Coffee Martini}}
        \label{fig:pd:coffee_martini}
    \end{subfigure}
    \hfill
    \begin{subfigure}[b]{0.48\textwidth}
        \centering
        \includegraphics[width=\textwidth]{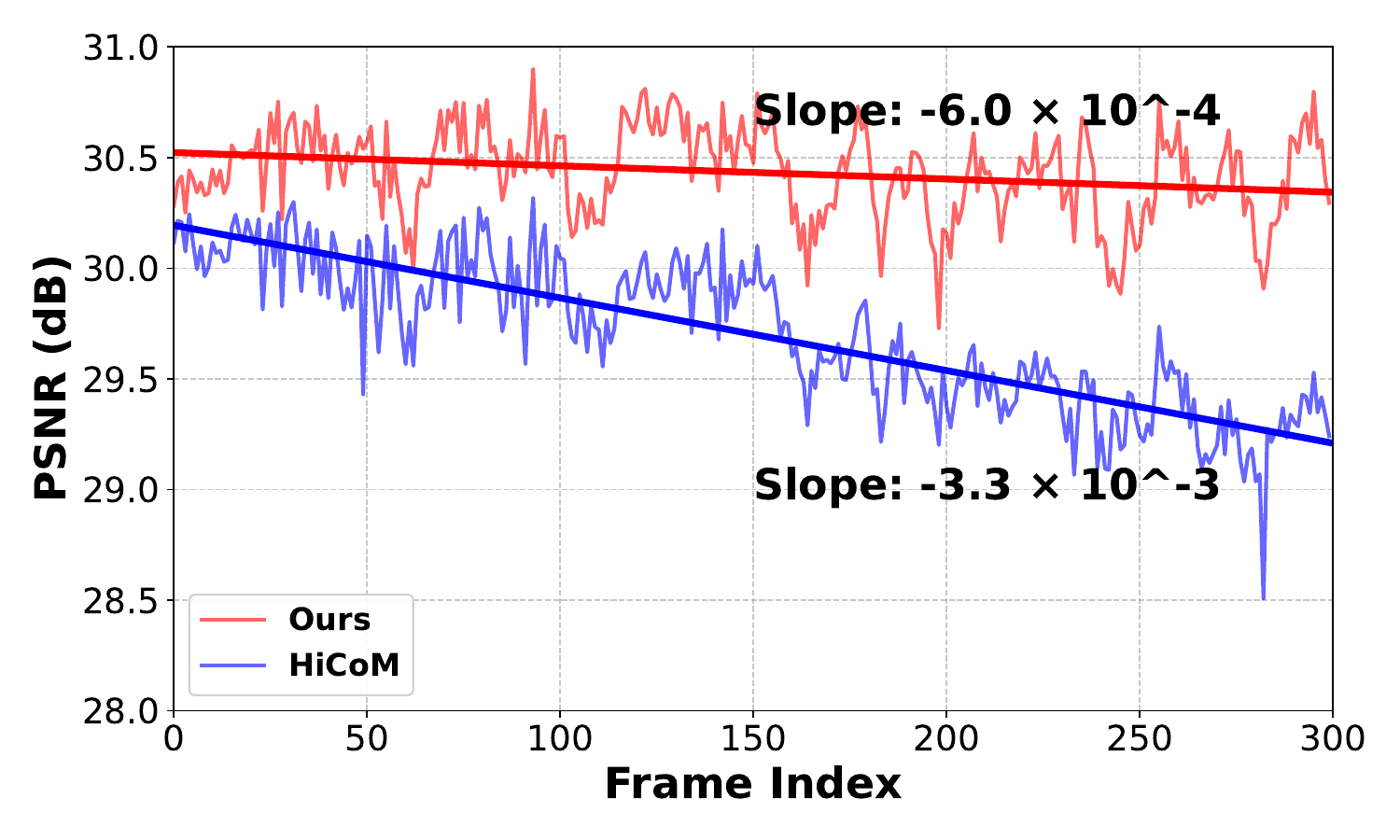}
        \caption{\textit{Flame Salmon}}
        \label{fig:pd:flame_salmon}
    \end{subfigure}
    \caption{The PSNR Trend Comparison between Ours and HiCoM~\cite{hicom} across different scenarios.}
    \label{fig:performance_drop}
\end{figure}

\subsection{Dataset}
\label{sec:experiment:dataset}
\textbf{The Nerual 3D Video Dataset (N3DV)}~\cite{n3dv}~comprises 6 indoor dynamic scenes. Each scene contains 18 to 21 free-viewpoint videos, with each video spanning 300 frames captured at 30 FPS and a resolution of $2704\times2028$.
%N3DV数据集由6个室内的动态场景组成。每个场景包含21个自由视点视频，一共300帧，帧率为30FPS，分辨率为2704x2028。

\textbf{Meeting Room Dataset}~\cite{streamrf}~comprises 3 dynamic scenes across diverse real-world scenarios. Each scene contains 13 free-viewpoint videos, with each video spanning 300 frames captured at 30 FPS and a resolution of $1280\times720$.
% Meetroom数据集由3个不同情境下的动态场景组成。每个场景包含13个自由视点视频，一共300帧，帧率为30FPS，分辨率为1280x720。

\textbf{PanopticSports Dataset}~\cite{dynamic3dgs}, derived from sports sequences in the Panoptic Studio dataset~\cite{panopticstudio}, includes 6 sports-oriented scenarios. Each scenario consists of 31 free-viewpoint videos, each containing 150 frames recorded at 30 FPS and a resolution of $640\times360$.
% PanopticSports数据集是由Panoptic Studio数据集中的6个运动序列组成的。每个场景包含31个自由视点视频，一共150帧，帧率为30FPS，分辨率为640x360。

\textbf{Technicolor Dataset}~\cite{technicolor} consists of video recordings captured by a $4 \times 4$ camera rigs. Each video has a spatial resolution of $2048 \times 1088$ and a frame rate of 30 FPS. Following the Ex4DGS~\cite{fedgs}, evaluation is performed on five distinct scenes (\textit{Birthday, Fabien, Painter, Theater, and Trains}) using their original full resolution.
% 跟随Ex4DGS的方法，我们在选取五个序列进行测试。

\subsection{Implementation Details}
\label{sec:experiment:imple}
% 我们使用默认超参数配置的3DGS搭配$\lambda_N=0.01$分别对三个训练集进行10000步到15000步的首帧训练。随后，在层级化过程中，我们将最细密的一层的锚点高斯总数为总高斯元数量的1/32，随后每层锚地高斯的数量为前一层的1/3。在后续每帧更新变形场参数时，我们会将每个数值乘以0.75以保留前一帧的运动趋势。对于三个数据集，我们每帧首先使用100步进行变形的训练，随后对N3DV和Meetroom会设置额外的100步对PanopticSports会设置额外的400步进行高斯元进行高斯元的微调与删减。我们使用HiCoM的官方代码以及其在文章中声明的训练参数作为我们的主要对比方法。所有的实验都在RTX4090 GPU上进行训练，并且取多次实验的平均值以保证结果的可靠性。更多的细节在附录中提供。
Our framework initializes first-frame reconstruction using 3DGS~\cite{3dgs} with the maximum degree of spherical harmonics (SH) to 1 and a noise-injected coefficient $\lambda_{noise}=0.01$, training on three datasets for 10,000 to 15,000 steps. During the construction of hierarchical motion representation, we set the finest-level Anchor Gaussians to 1/24 of the total Gaussian primitive count, with each subsequent layer containing one-third the anchors of its predecessor. 

For per-frame optimization, all datasets undergo an initial 100 steps for deformation training. Subsequently, N3DV and Meet Room datasets receive an additional 100 refinement steps for Gaussian densification and pruning, while PanopticSports employs 400 steps to address its higher motion complexity. All experiments are conducted on an NVIDIA RTX 4090 GPU, with results averaged over 3 independent runs to ensure statistical reliability. Extended implementation details are provided in the \textbf{Appendix~\ref{appenA:sec:imple}}.

\begin{figure}[!t]
  \centering
  \includegraphics[width=0.98\textwidth]{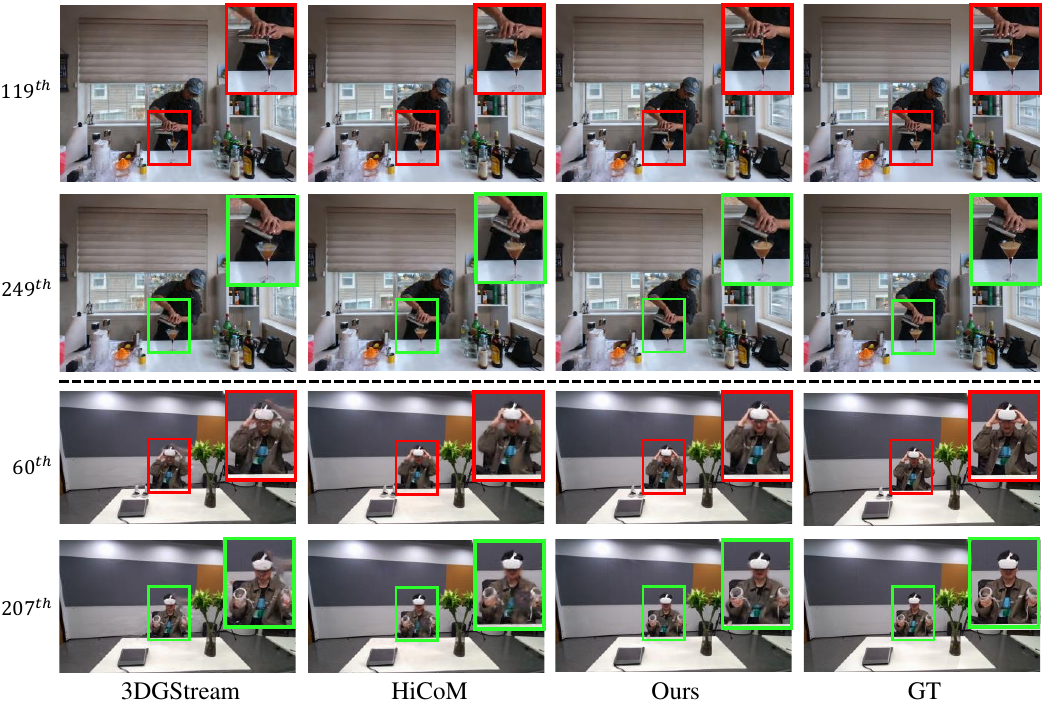}
  \caption{\textbf{Qualitative results} on the \textit{coffee martini} scene in the N3DV Dataset. To ensure a fair comparison, we retrained their official code with the same initial sparse points. }
  \label{fig:quality_main}
\end{figure}
\begin{figure}[t]
  \centering
  \includegraphics[width=1\textwidth]{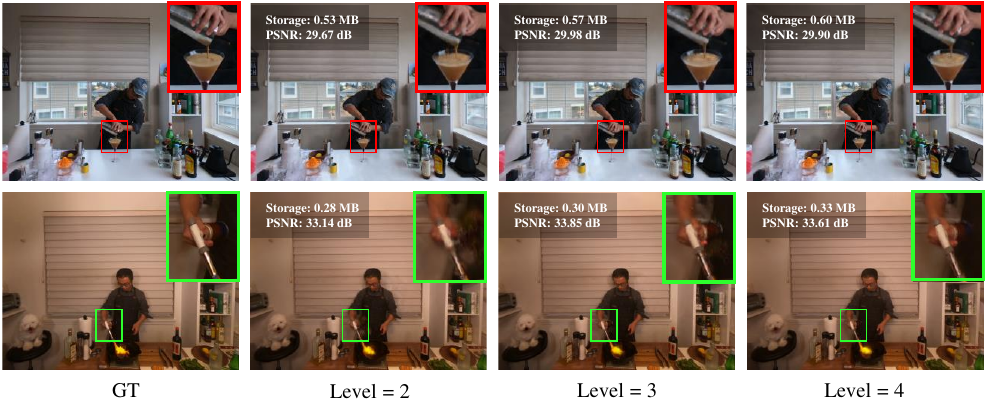}
  \caption{Qualitative results of our ReCon-GS under different hierarchical levels}
  \label{fig:quality_hier_depth}
\end{figure}
\begin{table}[t]
    \caption{
        \textbf{Ablation on key components of our ReCon-GS framework}. 
    }
    \label{tab:ablation_module}
    \centering
    \footnotesize
    \setlength{\tabcolsep}{2pt}
    \begin{tabular}{l|ccccc}
    \toprule
     Method & PSNR (dB)$\uparrow$ & SSIM $\uparrow$ & Storage (MB)$\downarrow$ & Train (sec)$\downarrow$ & Render (FPS) $\uparrow$ \\
     \midrule
      \small{w/o Hierarchical Motion Representation} & 31.43 & 0.9525 & \textbf{0.32} & \textbf{6.37} & 260\\
      \small{w/o Dynamic Hierarchy Reconfiguration}  & 32.00 & 0.9521 & 0.45 & 6.45 & 248\\
      \small{w/o View-based Densification}           & 32.15 & 0.9559 & 0.40 & 6.52 & \textbf{287}\\
      Ours~(full)                            & \textbf{32.66} & \textbf{0.9571} & 0.44 & 6.44 & 250\\
     \bottomrule
    \end{tabular}
\end{table}
\begin{table}[!ht]
    \caption{
        \textbf{Ablation on hierarchy depth of our ReCon-GS framework}. 
    }
    \label{tab:ablation_level}
    \footnotesize
    \centering
    \setlength{\tabcolsep}{4pt}
    \begin{tabular}{ccccccc}
    \toprule
\multirow{2.5}{*}{\begin{tabular}[c]{@{}c@{}}\small{Hierarchy}\\[-0.2ex]\small{Depth}\end{tabular}} &
\multicolumn{3}{c}{N3DV} & 
\multicolumn{3}{c}{Meet Room} \\
     \cmidrule(r){2-4} \cmidrule{5-7} 
     & PSNR (dB)~$\uparrow$ & Storage (MB)~$\downarrow$ & Train (sec)~$\downarrow$ & PSNR (dB)~$\uparrow$ & Storage (MB)~$\downarrow$ & Train (sec)~$\downarrow$\\

     \midrule
     2 & 32.422 & \textbf{0.40} & \textbf{6.44} & 30.20 & \textbf{0.28} & 3.87\\
     3 & \textbf{32.662} & 0.44 & 6.49 & 30.84 & 0.30 & \textbf{3.86}\\
     4 & 32.658 & 0.46 & 6.54 & \textbf{31.01} & 0.32 & 4.01\\
     \bottomrule
    \end{tabular}
\end{table}
\subsection{Experiment Results}
\label{sec:experiment:results}
\textbf{Quantitive comparisons.}
% 我们在三个数据集上使用了多个评价指标综合的对比了我们的方法与现有的在线和离线的sota方法，评价指标包括：PSNR，SSIM，LIPIS，Storage，Training Time以及Rendering Spped。我们的Storage和Rendering Speed是取得全部帧的平均值以保证数值的有效性。从表格1以及表格2中，我们的方法在所有的online方法中在所有指标上都是领先的。在所有数据集上，我们的方法在渲染质量和训练时间上实现了SOTA的性能。并且，在N3DV、MeetRoom以及Technicolor数据集上，我们的方法均比目前SOTA的流式方法的渲染质量高大于0.5dB。在PanopticSports数据集上，我们的方法则在SOTA的渲染质量下较之前的方法减小了约60%的存储。对比Offline方法，ReCon-GS在存储上略高，主要是归咎于离线方法在隐式运动表示上的天然存储优势。
% 而在渲染时间和训练时间上，我们的方法都比目前的sota方法要更好。此外，根据图X可知，在保证渲染质量的情况下，我们的方法可以比现有的在线sota方法实现超过50%的存储下降。
We conducted a comprehensively evaluation of our method against existing state-of-the-art (SOTA) online and offline approaches across three datasets using multiple quantitative metrics, including PSNR, SSIM, LPIPS, storage consumption, training time, and rendering speed. To ensure statistical validity, storage and rendering speed measurements were calculated as averaged values across all video frames. As demonstrated in Table~\ref{tab:benchmark1}~, Table~\ref{tab:benchmark2} and Table~\ref{tab:technicolor_benchmark}, our method achieves SOTA performance in rendering quality and training efficiency compared to online baselines.
Specifically, our method outperforms the recent SOTA streaming methods in rendering quality by \textbf{more than 0.5dB} on the N3DV~\cite{n3dv}, MeetRoom~\cite{streamrf}, and Technicolor~\cite{technicolor} datasets. This improvement underscores the effectiveness of our approach in achieving superior visual fidelity in dynamic scene reconstructions. In the case of the PanopticSports~\cite{dynamic3dgs} dataset, our method achieves the best rendering quality while inducing a substantial reduction in storage requirements, decreasing storage by approximately 60\% compared to previous streaming methods. In comparison to offline methods, ReCon-GS requires slightly more storage, primarily due to the inherent storage advantages of offline methods in implicit motion representation.
% While our method demonstrates storage superiority over online counterparts, it shows relatively higher storage consumption compared to some offline methods. This discrepancy can be attributed to the inherent storage benefits of offline methods derived from their implicit motion representations.

Furthermore, as evidenced in Figure~\ref{fig:performance_drop}, ReCon-GS effectively alleviates temporal performance degradation observed in recent SOTA method. Through the Dynamic Hierarchy Reconfiguration Strategy, ReCon-GS successfully circumvents the degradation in motion representation capability of Anchor Gaussians. More quantitive evaluations are provided in the \textbf{Appendix~\ref{appendix:quantitative}}.

\textbf{Qualitative comparisons.}
% 尽管我们的ReCon-GS框架主要是优化动态场景重建的效率，但是表格1和表格2也说明我们的方法在渲染质量上的先进性。如图X所示，较于HiCoM以及3DGStream，我们的方法在多个数据集上实现了对场景中的动态细节更加完整、平滑的捕捉。并且，在相对长的时间间隔时，我们的方法可以实现更加好的动态重建。更多的主观质量测试在Appendix中提供。
While our ReCon-GS framework primarily focuses on optimizing the efficiency of dynamic scene reconstruction, Table~\ref{tab:benchmark1} and Table~\ref{tab:benchmark2} substantiate the superiority of our method in rendering quality. As illustrated in Figure~\ref{fig:quality_main}, compared to HiCoM~\cite{hicom} and 3DGStream~\cite{3dgstream}, our approach achieves more complete and temporally consistent reconstruction of dynamic scene details across multiple datasets. Moreover, the proposed method demonstrates more enhanced dynamic reconstruction capabilities when handling prolonged temporal intervals.
We have also provided an additional subjective quality evaluations in \textbf{Appendix~\ref{appendix:qualitative}}.
\subsection{Ablation Study}
\label{sec:experiment:ablation}
% 从表格3可以看出，我们对ReCon-GS的主要策略进行了消融实验，包括是否使用Adaptively Hierarchical Motion Presentation，是否对首帧使用AWGN，是否进行重-层级化以及是否进行基于视角的密集化过程。结果可以看出，对首帧使用AWGN对高斯元位置进行干扰以及重-层级化的策略有效的提升了ReCon-GS的性能，代表这些策略有效的加强了ReCon-GS对于高斯元的时-空位置捕捉。虽然基于视角的密集化会导致ReCon-GS的每帧训练时间增加，但是密集化过程有效的加强了ReCon-GS对于场景高频细节的表示。这意味着只对场景进行基础高斯元的变形是不足以实现对动态场景的完整表示的。
% 此外，从表格4可以看出，我们对层级化的数量也进行消融实验，这主要是因为我们希望借助ReCon-GS探究运动表达层级个数的高效性，实验证明当层级数量为3时，ReCon-GS同时实现了场景的高保真表示与存储的权衡。
As evidenced in Table~\ref{tab:ablation_module}, we conducted ablation studies on key strategies of ReCon-GS, including Adaptively Hierarchical Motion Presentation paradigm, Dynamic Hierarchy Reconfiguration strategy and view-based densification process. 
The experimental results demonstrate that the Adaptively Hierarchical Motion Presentation paradigm, despite introducing some memory overhead, significantly enhances the performance of ReCon-GS, proving the paradigm's effective motion representation capabilities. 
Furthermore, the implementation of Dynamic Hierarchy Reconfiguration strategy further improves ReCon-GS's rendering quality, indicating enhanced spatiotemporal localization accuracy of Gaussian primitives and ensured temporal motion consistency. 
While the view-based densification process increases per-frame training time, it substantially improves the representation of high-frequency scene details. This suggests that merely deforming base Gaussian primitives is insufficient for comprehensive dynamic scene modeling.

Furthermore, Table~\ref{tab:ablation_level} presents ablation experiments on hierarchy depth, motivated by our investigation into motion representation efficiency through ReCon-GS. As shown in Figure~\ref{fig:quality_hier_depth}, experimental findings reveal that a 3-level hierarchy optimally balances high-fidelity scene reconstruction with storage efficiency. This configuration enables ReCon-GS to achieve superior fidelity-storage trade-offs compared to other hierarchical configurations. See \textbf{Appendix~\ref{appendix:ablation}} for more ablation studies.

\section{Conclusion}
\label{sec::conclusion}
\vspace{-5pt}
% 这篇文章引入了\textbf{ReCon-GS}，一种面向多视点视频重建的存储感知框架。\textbf{ReCon-GS}提出了自适应分层运动表示框架从而实现了高斯元密度感知的多层解耦运动紧凑表示，这解决了现有方法因为基于空间均匀的运动表示而导致低效问题。此外我们提出了一种双-层级变形策略。通过时域帧间变形与空域帧内变形，\textbf{ReCon-GS}更好的捕捉了高斯元的时-空域位置，避免出现对于高动态运动以及新出现物体重建不好的问题。此外，得益于\textbf{ReCon-GS}的框架特性，我们将动态场景重建问题从以往的单一目标优化问题转换到存储感知的保真度优化问题，我们的方法能够根据应用场景需求动态平衡内存效率与渲染质量，极大地提升了3D动态重建方法的实用性。广泛的实验结果显示\textbf{ReCon-GS}实现了state-of-the-art的渲染质量以及训练时间的同时，保证了存储的高效性。对比现有sota方法，\textbf{ReCon-GS}在相当渲染质量的情况下可以实现超过50%的内存节省。
This paper introduces \textbf{ReCon-GS}, a novel storage-aware framework for high-fidelity multi-view video reconstruction. ReCon-GS achieves density-aware multi-layer decoupled compact motion modeling of Gaussian primitives through an Adaptive Hierarchical Motion Representation framework. Furthermore, we propose a Dual-level Deformation Strategy combining inter-frame temporal deformation and intra-frame spatial deformation, accurately capture spatio-temporal positions of Gaussian primitives while avoiding reconstruction failures in high-dynamic motions and emerging objects.
Notably, ReCon-GS innovatively transforms dynamic scene reconstruction from conventional single-objective optimization into a storage-aware fidelity optimization framework. This paradigm shift allows dynamic balancing between memory efficiency and rendering quality according to application requirements, significantly enhancing the practical applicability of 3D dynamic reconstruction method. Extensive experiments demonstrate that ReCon-GS achieves state-of-the-art performance in both rendering quality and training efficiency while maintaining exceptional storage effectiveness. Compared with existing SOTA methods, ReCon-GS attains over 50\% memory reduction while preserving same rendering fidelity.

\textbf{Limitations.}
% 与之前的方法一样，我们的方法专注于多视点通用场景重建任务，并没有对单目数据集以及特定类型的重建任务进行优化。此外，流式框架的特性使得我们与其他方法一样通常面临着误差累计的问题。
While maintaining alignment with prior methodologies in focusing on multi-view general scene reconstruction tasks, our approach is not specifically optimized for monocular datasets or specialized reconstruction scenarios. Similar to existing streaming-based architectures, we face challenges in dynamic allocation/updating of Gaussians to handle real-time object emergence/dissolution, as abrupt scene changes require instant 3DGS structural adaptations.

\newpage
{
\small
\bibliographystyle{unsrt}

}

\newpage

\appendix

\section*{Appendix}
\section{More Implementation Details}
\label{appenA:sec:imple}
\vspace{-6pt}
% 我们的ReCon-GS是基于3DGS的开源代码基础进行开发的。在首帧的训练时，我们跟随3DGStream的首帧训练方法，将球谐系数设置为1。此外，对于所有数据集，3DGS的Densication会停止在5000 iterations，这极大地缓解了流式3DGS在首帧过拟合的问题。此外，我们采取了4DGS中不会对重置不透明度的策略，这极大地消除了那些大的高斯元。此外，对于$\lambda_{noise}$的选择，我们跟随了HiCoM的首帧设置。除此之外，首帧训练中其他3DGS超参数我们都使用了默认3DGS的参数。对MeetRoom数据集，我们对第一帧进行了10000步的训练。对于N3DV以及PanopticSports数据集，我们对第一帧进行了15000步的训练。

% 在后续帧的训练中，不同于HiCoM在PanopticSports使用了250步对变形场进行训练，这与N3DV与MeetRoom中Stage 1只训练100步不同，我们对所有的数据集都进行了100步的变形场训练。而在Stage2，也就是每帧分别对变形后的高斯元进行refine的过程，由于PanopticSports在不同时间下物体的表面形变相较于N3DV以及Meet Room数据集更加剧烈，因此我们对PanopticSports数据集进行了400步的密集化训练，对N3DV和MeetRoom数据集进行了100步的Stage 2训练。其中所有的3DGS相关超参数都与首帧一致。
We implement ReCon-GS upon the open-source codebase of 3D Gaussian Splatting (3DGS)~\cite{3dgs}. During initial frame training, we set the spherical harmonic coefficients (SH) to degree 1 to ensure the storage compactness. For all datasets, we terminate the densification process of 3DGS at 5,000 iterations, which significantly alleviates the overfitting issue in streaming 3DGS during initial frame training. Additionally, we adopt the opacity preservation strategy from 4DGS to effectively eliminate oversized Gaussian primitives. The hyperparameter $\lambda_{noise}$ follows the first-frame configuration from HiCoM~\cite{hicom}, while other 3DGS parameters maintain their default settings in 3DGS. Specifically, we train the first frame for 10,000 iterations on the Meet Room dataset, and extend this to 15,000 iterations for both N3DV and PanopticSports datasets.

For subsequent frame optimization, our approach differs from HiCoM's implementation which uses 250 iterations for deformation field training on Panoptic Sports versus 100 iterations on N3DV and Meet Room datasets. We uniformly apply 100 iterations for deformation field training across all datasets due to our superior ability in capturing the motion of Gaussian primitives. In Phase 2, the refinement of deformed Gaussian primitives, considering the more pronounced surface deformation in PanopticSports dataset compared to other datasets, we implement 400 iterations for intensive refinement on PanopticSports dataset versus 100 iterations for N3DV and MeetRoom datasets. All 3DGS-related hyperparameters remain consistent with those used in initial frame training to ensure methodological coherence.

\vspace{-6pt}
\section{More Ablation Studies}
\label{appendix:ablation}
\vspace{-6pt}
\begin{table}[b]
    \caption{
        \textbf{Ablation on training steps of our ReCon-GS framework}. 
    }
    \label{tab:ablation_training_step}
    \centering
    \setlength{\tabcolsep}{5pt}
    \begin{tabular}{cc|ccccc}
    \toprule
     Phase 1 & Phase 2 & PSNR (dB)$\uparrow$ & SSIM $\uparrow$ & Storage (MB)$\downarrow$ & Train (sec)$\downarrow$ & Rendering (FPS) $\uparrow$ \\
     \midrule
     50  & \multirow{2}{*}{100} & 32.40 & 0.9560 & \textbf{0.44} & \textbf{5.20} & 248\\
     150 &                      & \textbf{32.76} & 0.9568 & \textbf{0.44} & 7.66 & 249\\
     \midrule
     \multirow{2}{*}{100} & 50  & 32.56 & 0.9564 & 0.46 & 5.22 & 246\\
                          & 150 & 32.64 & 0.9568 & 0.49 & 7.70 & 249\\
     \midrule
     100 & 100 & 32.66 & \textbf{0.9571} & \textbf{0.44} & 6.44 & \textbf{250} \\
     \bottomrule
    \end{tabular}
\end{table}
% 由于ReCon-GS是一种面向高保真实时渲染的动态场景重建的新型存储感知框架，我们自适应分层运动表示结构每层的锚点高斯数量决定了ReCon-GS对于存储效率与渲染质量的权衡。而，每个Stage的训练步长则是训练时间的决定因素。表格X探究了在常用的各阶段步长设置对于ReCon-GS的影响。# 需要进一步补充对实验结果的分析
As ReCon-GS represents a novel storage-aware framework designed for high-fidelity real-time rendering in dynamic scene reconstruction, the quantity of Anchor Gaussians per layer in our adaptive hierarchical motion representation structure fundamentally governs the critical trade-off between storage efficiency and rendering quality. Concurrently, the iteration counts allocated to each training stage in subsequent frame training directly determine the training time of our optimization process. 
%Table~\ref{tab:ablation_training_step} systematically investigates the impacts of different iteration configurations across training phases on ReCon-GS performance.\colorbox{yellow}{More Analysis according to ablation results need to be drafted....... TBD}

% 表格X探究了在常用的各阶段步长设置对于ReCon-GS的影响。结果显示，延长Stage 1的变形场训练过程可以有效的增长渲染质量，然而这种训练对于训练时间的影响也是极大的。而Stage 2的训练补偿则是在100步时呈现了最优的性能以及最佳的存储质量。从表格～\ref{}和表格~\ref{}结果来看，密集化的步骤有效的修正了persistent discrepancies between learned motion and scene dynamics，但是其对于整体渲染质量的影响是小于变形场的。
Table~\ref{tab:ablation_training_step} investigates the impacts of different iteration configurations across training phases on ReCon-GS performance. The results indicate that prolonging the deformation field training process in Phase 1 can effectively improve rendering quality, however, this training configuration significantly increases computational time. The view-based densification process in Phase 2 demonstrates optimal performance and best storage quality at 100 steps. As evidenced by Table~\ref{tab:ablation_module} and Table~\ref{tab:ablation_training_step}, view-based densification effectively address persistent discrepancies between learned motion and scene dynamics, though their impact on overall rendering quality remains less substantial compared to the training of adaptively hierarchical deformation.

% 我们还针对不同首帧训练情况下ReCon-GS的性能进行了测试。从表格X中，我们可以看到， At 2,500 iterations, suboptimal initial reconstruction quality and sparse scene representation result in the steepest performance degradation slope. When trained for 5,000 iterations, quality degradation is substantially mitigated. Extending to 10,000 iterations yields converged initial frame training, further optimizing quality decay while achieving state-of-the-art (SOTA) performance. Responding to your insight, we will integrate this analysis in the future revised version to demonstrate the critical impact of initial frame quality on streaming frameworks while further validating ReCon-GS's superior robustness against error accumulation.
We also performed an ablation study on the performance of ReCon-GS, varying the training duration of the initial frame (Frame 0). The detailed results is shown in Table~\ref{tab:ablation_init_frame_training}. At 2,500 iterations, the suboptimal quality and sparse representation of the initial reconstruction lead to the steepest performance degradation slope. Increasing the training to 5,000 iterations substantially mitigates this quality degradation. Extending the training to 10,000 iterations yields a converged initial frame, which further minimizes quality decay while achieving state-of-the-art performance. Importantly, with the converged initial frame, ReCon-GS requires only a minimal number of incremental Gaussians per frame to ensure high-quality reconstruction. This result strongly validates the superior robustness of ReCon-GS against error accumulation and underscores the critical impact of initial frame quality in streaming frameworks.
\begin{table}[t]
    \caption{
        \textbf{Ablation on the Training Steps of Frame 0}. The Init. 3DG refers to the initial 3D Gaussians constructed from the Frame 0. The Incr. 3DG denotes the incremental Gaussians added in Phase 2 for all following frames.
    }
    \label{tab:ablation_init_frame_training}
    \centering
    \setlength{\tabcolsep}{4pt}
    \begin{tabular}{c|cccccc}
    \toprule
    \multirow{2.5}{*}{Training Steps} & PSNR & Frame 0 PSNR & Slope &  Storge & Init. 3DG Storage & Incr. 3DG Storage \\
    & (dB)~$\uparrow$ & (dB)~$\uparrow$ & $\downarrow$ & (MB)~$\downarrow$ & (MB)~$\downarrow$ & (MB)~$\downarrow$ \\
     \midrule
     2500  & 31.03 & 31.59 & 0.0077 & 0.46 & 8.96 & 42.17  \\
     5000  & 32.24 & 32.51 & 0.0029 & 0.48 & 11.14 & 11.57 \\
     10000 & 32.63 & 32.75 & 0.0016 & 0.46 & 11.12 & 4.11  \\ 
     \bottomrule
    \end{tabular}
\vspace{-6pt}
\end{table}

\vspace{-6pt}
\section{More Detailed Results}
\vspace{-4pt}
\subsection{Quantitative Results}
\vspace{-4pt}
\label{appendix:quantitative}
\subsubsection{Initial 3DGS Comparison}
\vspace{-4pt}
% 对于流式3DGS重建框架，首帧训练对后续帧的重建质量起到了决定性作用。在表格X中，我们展示了在三个数据集下平均的首帧训练质量以及高斯元数量即存储的开销。由于我们的ReCon-GS与HiCoM都采取了向首帧的训练中向高斯元位置注入噪声的策略，相较于3DGStream，我们的方法具有更高的渲染质量。同时，因为ReCon-GS的首帧训练时将球谐系数（SH）设置为1，因此我们的方法相较于HiCoM也具有更有优势的存储效率。
\begin{table*}[t]
\centering  
\footnotesize
\caption{\textbf{First-frame reconstruction results across various scenario configurations.} 3DGStream uses standard 3DGS training, Our Method and HiCoM use standard 3DGS training that incorporates with positional noise injection.}  
\label{tab:first_frame_performance}  
\setlength{\tabcolsep}{1.4pt}
\begin{tabular}{l cccc cccc cccc}  
    \toprule  
    \multirow{3.5}{*}{Method} &  
    \multicolumn{4}{c}{Coffee Martini}&\multicolumn{4}{c}{Flame Steak}&\multicolumn{4}{c}{Sear Steak} \\
    \cmidrule(r){2-5} \cmidrule{6-9} \cmidrule(r){10-13}
    & PSNR & GS Num & Storage & Train & PSNR & GS Num & Storage & Train & PSNR & GS Num & Storage & Train \\
    & (dB)$\uparrow$ & (k)$\downarrow$  & (MB)$\downarrow$ & (Sec)$\downarrow$ & (dB)$\uparrow$ & (k)$\downarrow$  & (MB)$\downarrow$ & (Sec)$\downarrow$ & (dB)$\uparrow$ & (k)$\downarrow$ & (MB)$\downarrow$ & (Sec)$\downarrow$ \\
    \midrule
    3DGStream    & 27.62 & 416 & 7.60  & 340 & 33.85 & 208 & 7.60  & 304 & 29.25 & 214 & 7.60  & 317 \\
    HiCoM            & 29.73 & 348 & 83.00 & 368 & 34.27 & 187 & 44.55 & 334 & 33.96 & 191 & 45.60 & 317 \\
    \textbf{Ours}                 & 29.67 & 353 & 16.17 & 354 & 31.20 & 189 & 8.65  & 324 & 33.90 & 191 & 8.76  & 338  \\
    \bottomrule  
\end{tabular}  
\end{table*}
For streaming 3DGS reconstruction frameworks, the initial frame training plays a pivotal role in determining the reconstruction quality of subsequent frames. Table~\ref{tab:first_frame_performance} presents comparative metrics of average initial frame training quality and corresponding storage consumption across three benchmark datasets. While both ReCon-GS and HiCoM implement positional noise injection during initial Gaussian optimization – a strategic commonality that differentiates them from 3DGStream – our framework demonstrates superior rendering fidelity. This configuration, combined with our first-frame spherical harmonic (SH) coefficient restriction to degree 1, enables ReCon-GS to achieve more compact Gaussian representations compared to HiCoM's SH degree 3 implementation, significantly enhancing storage efficiency without compromising visual quality.

% 添加关于训练时间的描述

% \subsubsection{Additional Quantitative Results}
% %我们提供了在Technicolor数据集上额外的定量实验。Technicolor数据集是XXXXXX。从表格X我们可以看到，ReCon-GS在Technicolor上在所有比较metric上实现了SOTA。需要注意的是，相较于最新的流式方法， ReCon-GS delivers approximately 0.5 dB higher rendering quality while reducing storage by over 40% and significantly accelerating rendering speed. Against offline methods, ReCon-GS attains SOTA rendering quality while achieving more than 25% storage reduction.
% We present additional quantitative evaluations on the Technicolor dataset~\cite{technicolor}. Techicolor dataset is a multiview dataset captured with a time-synchronized $4 \times 4$ camera rig. The results, summarized in Table~\ref{tab:technicolor_benchmark}, demonstrate that ReCon-GS establishes new state-of-the-art performance across all evaluated metrics. Specifically, compared to the latest streaming methods, ReCon-GS delivers approximately 0.5 dB higher rendering quality while reducing storage by over 40\% and significantly accelerating rendering speed. Against offline methods, ReCon-GS attains SOTA rendering quality while achieving more than 25\% storage reduction.
% \input{Table/technicolor_supple}

\begin{table*}[!t]
\centering  
\footnotesize
\caption{\textbf{Per-scene quantitative results on the N3DV dataset.} The method with $^\dag$ is reproduced by us through official code in the same experimental environment.}  
\label{tab:dynerf_scenes}  
\setlength{\tabcolsep}{2.0pt}
\begin{tabular}{l l ccc ccc ccc}  
    \toprule  
    \multirow{2.5}{*}{Category} & \multirow{2.5}{*}{Method} &  
    \multicolumn{3}{c}{Coffee Martini}&\multicolumn{3}{c}{Cook Spinach}&\multicolumn{3}{c}{Cut Roasted Beef} \\
    \cmidrule(r){3-5} \cmidrule{6-8} \cmidrule(r){9-11}
    & & PSNR$\uparrow$ & SSIM$\uparrow$ & Storage$\downarrow$ & PSNR$\uparrow$ & SSIM$\uparrow$ & Storage$\downarrow$ & PSNR$\uparrow$ & SSIM$\uparrow$ & Storage$\downarrow$\\
    \midrule
    \multirow{2}{*}{Offline} 
    & 4DGS$^\dag$~\cite{4d-gs}    & \cellsecond 30.10 & \cellsecond 0.935 & \cellbest 0.30 & 31.15 & 0.958 & \cellbest 0.30 & 32.60 & 0.951 & \cellbest 0.30 \\
    & SaRO-GS~\cite{sarogs}       & 28.96 & - & 1.0 & 33.19 & - & 1.0 & 33.91 & - & 1.0 \\
    \midrule
    \multirow{5}{*}{Online} 
    & 3DGStream$^\dag$~\cite{3dgstream}  & 27.75 & 0.917 & 7.85 & 32.22 & 0.957 & 7.79 & 32.67 & 0.957 & 7.76 \\
    & 4DGC~\cite{4dgc}            & 27.98 & - & \cellsecond 0.58 & 32.81 & - & 0.44 & 33.03 & - & 0.47 \\
    & QUEEN-l~\cite{queen}        & 28.38 & 0.915 & 1.17 & 33.40 & 0.956 & 0.59 & \cellbest 34.01 & 0.959 & 0.57 \\ 
    & HiCoM$^\dag$~\cite{hicom}   & 28.67 & 0.925 & 0.82 & \cellbest 33.73 & \cellbest 0.962 & 0.63 & 32.73 & 0.963 & 0.60 \\
    & \textbf{Ours}               & \cellbest 30.14 & \cellbest 0.938 & 0.64 & \cellsecond 33.54 & \cellsecond 0.961 & \cellsecond 0.35 &  \cellsecond 33.92 & \cellbest 0.966 & \cellsecond 0.36 \\
    \toprule  
    & & \multicolumn{3}{c}{Flame Salmon} & \multicolumn{3}{c}{Flame Steak} & \multicolumn{3}{c}{Sear Steak} \\ 
    \cmidrule(r){3-5} \cmidrule{6-8} \cmidrule(l){9-11}
    & & PSNR$\uparrow$ & SSIM$\uparrow$ & Storage$\downarrow$ & PSNR$\uparrow$ & SSIM$\uparrow$ & Storage$\downarrow$ & PSNR$\uparrow$ & SSIM$\uparrow$ & Storage$\downarrow$\\
    \midrule
    \multirow{2}{*}{Offline}
    & 4DGS$^\dag$~\cite{4d-gs}    & 30.21 & \cellsecond 0.934 & \cellbest 0.30 & 33.49 & 0.961 & \cellbest 0.30 & 30.64 & 0.961 & \cellbest 0.30 \\
    & SaRO-GS~\cite{sarogs}       & 29.14 & - & 1.0 &  33.83 & - & 1.0 &  33.89 & - & 1.0 \\
    \midrule
    \multirow{5}{*}{Online} 
    & 3DGStream$^\dag$~\cite{3dgstream} & 28.61 & 0.924 & 7.83 & 33.47 & \cellsecond 0.966 & 7.79 & 33.39 & 0.965 & 7.76 \\
    & 4DGC~\cite{4dgc}            & 28.49 & - & \cellsecond 0.51 & 33.58 & - & 0.44 & 33.60 & - & 0.50 \\
    & QUEEN-l~\cite{queen}              & 29.25 & 0.923 & 1.00 & \cellbest 34.17 & 0.962 & 0.59 & \cellbest 33.93 & 0.962 & 0.56 \\
    & HiCoM$^\dag$~\cite{hicom}         & \cellsecond 29.70 & 0.932 & 0.81 & 33.92 & \cellbest 0.969 & 0.60 & 33.71 & \cellbest 0.968 & 0.60 \\
    & \textbf{Ours}                     & \cellbest 30.43 & \cellbest 0.938 & 0.61 & \cellsecond 34.00 & \cellbest 0.969 & \cellsecond 0.33 & \cellsecond 33.91 & \cellsecond 0.966 & \cellsecond 0.35 \\
    \bottomrule  
\end{tabular}  
\end{table*}
\begin{table*}[!t]
\centering  
\footnotesize
\caption{\textbf{Per-scene quantitative results on the Meet Room dataset.} The method with $^\dag$ is reproduced by us through official code in the same experimental environment.}  
\label{tab:meetroom_scenes}  
\setlength{\tabcolsep}{2.0pt}
\begin{tabular}{l l ccc ccc ccc}  
    \toprule  
    \multirow{2.5}{*}{Category} & \multirow{2.5}{*}{Method} &  
    \multicolumn{3}{c}{Discussion}&\multicolumn{3}{c}{Trimming}&\multicolumn{3}{c}{Vrheadset} \\
    \cmidrule(r){3-5} \cmidrule{6-8} \cmidrule(r){9-11}
    & & PSNR$\uparrow$ & SSIM$\uparrow$ & Storage$\downarrow$ & PSNR$\uparrow$ & SSIM$\uparrow$ & Storage$\downarrow$ & PSNR$\uparrow$ & SSIM$\uparrow$ & Storage$\downarrow$\\
    \midrule
    \multirow{3}{*}{Online} 
    & 3DGStream$^\dag$~\cite{3dgstream}    & \cellsecond 30.06 & \cellsecond 0.945 & 4.10 & 28.59 & 0.939 & 4.10 & 29.25 & 0.941 & 4.10 \\
    & HiCoM$^\dag$~\cite{hicom}            & 29.39 & 0.941 & \cellsecond 0.48 & \cellsecond 29.67 & \cellsecond 0.945 & \cellsecond 0.37 & \cellsecond 29.65 & \cellsecond 0.947 & \cellsecond 0.30 \\
    & \textbf{Ours}                        & \cellbest 30.67 & \cellbest 0.955 & \cellbest 0.35 & \cellbest 31.20 & \cellbest 0.956 & \cellbest 0.28 & \cellbest 30.64 & \cellbest 0.953 & \cellbest 0.30  \\
    \bottomrule  
\end{tabular}  
\end{table*}
\begin{table*}[!t]
\centering  
\footnotesize
\caption{\textbf{Per-scene quantitative results on the PanopticSports dataset.} The method with $^\dag$ is reproduced by us through official code in the same experimental environment.}  
\label{tab:panoptic_scenes}  
\setlength{\tabcolsep}{1.4pt}
\begin{tabular}{l l ccc ccc ccc}  
    \toprule  
    \multirow{2.5}{*}{Category} & \multirow{2.5}{*}{Method} &  
    \multicolumn{3}{c}{Basketball}&\multicolumn{3}{c}{Boxes}&\multicolumn{3}{c}{Football} \\
    \cmidrule(r){3-5} \cmidrule{6-8} \cmidrule(r){9-11}
    & & PSNR$\uparrow$ & SSIM$\uparrow$ & Storage$\downarrow$ & PSNR$\uparrow$ & SSIM$\uparrow$ & Storage$\downarrow$ & PSNR$\uparrow$ & SSIM$\uparrow$ & Storage$\downarrow$\\
    \midrule
    \multirow{1}{*}{Offline}
    & 4DGS$^\dag$~\cite{4d-gs}         & 26.99 & 0.912 & \cellbest 0.34 & 27.76 & 0.922 & \cellbest 0.32 & 27.20 & 0.922 & \cellbest 0.36  \\
    \midrule
    \multirow{4}{*}{Online} 
    & Dynamic3DGS~\cite{dynamic3dgs}   & 28.22 & 0.910 & \cellsecond 0.74 & \cellbest 29.46 &  0.910  & \cellsecond 0.74 & 28.49 &  0.910 & \cellsecond 0.74  \\
    & 3DGStream$^\dag$~\cite{3dgstream}& 20.08 & 0.815 & 8.10 & 23.76 & 0.861 & 8.10 & 22.80 & 0.856 & 8.10  \\
    & HiCoM$^\dag$~\cite{hicom}        & \cellsecond 28.49 & \cellsecond 0.915 & 2.06 & \cellsecond 29.03 & \cellsecond 0.922 & 2.05 & \cellsecond 29.64 & \cellsecond 0.926 & 2.01  \\
    & \textbf{Ours}                    & \cellbest 28.78 & \cellbest 0.921 & 0.81 & 28.93 & \cellbest 0.926 & 0.80 & \cellbest 30.01 & \cellbest 0.931 & 0.77  \\
    \toprule  
    & & \multicolumn{3}{c}{Juggle} & \multicolumn{3}{c}{Softball} & \multicolumn{3}{c}{Tennis} \\ 
    \cmidrule(r){3-5} \cmidrule{6-8} \cmidrule(l){9-11}
    & & PSNR$\uparrow$ & SSIM$\uparrow$ & Storage$\downarrow$ & PSNR$\uparrow$ & SSIM$\uparrow$ & Storage$\downarrow$ & PSNR$\uparrow$ & SSIM$\uparrow$ & Storage$\downarrow$\\
    \midrule
    \multirow{1}{*}{Offline}
    & 4DGS$^\dag$~\cite{4d-gs}                & 26.72 & 0.923 & \cellbest 0.32 & 26.88 & 0.923 & \cellbest 0.31 & 27.64 & 0.925 & \cellbest 0.31  \\
    \midrule
    \multirow{4}{*}{Online} 
    & Dynamic3DGS~\cite{dynamic3dgs}   & \cellsecond 29.48 & 0.920 & \cellsecond 0.74 & \cellsecond 28.43 & 0.910  & \cellsecond 0.74 & \cellsecond 28.11 & 0.910 & \cellsecond 0.74  \\
    & 3DGStream$^\dag$~\cite{3dgstream}& 26.25 & 0.891 & 8.10 & 22.22 & 0.858 & 8.10 & 22.99 & 9.862 & 8.10  \\
    & HiCoM$^\dag$~\cite{hicom}        & 29.28 & \cellsecond 0.932 & 2.02 & \cellbest 29.61 & \cellsecond 0.928 & 1.97 & \cellbest 28.97 & \cellsecond 0.925 & 2.05  \\
    & \textbf{Ours}                    & \cellbest 29.71 & \cellbest 0.938 & 0.79 & \cellbest 29.61 & \cellbest 0.933 & 0.77 & \cellbest 28.97 & \cellbest 0.929 & 0.79  \\
    \bottomrule  
\end{tabular}  
\end{table*}

\vspace{-4pt}
\subsubsection{Per-Scene Results}
\vspace{-4pt}
% 我们在表格X，X，X中展示了三个数据集的具体场景下ReCon-GS与最近几个SOTA的方法的更多对比结果。
Table~\ref{tab:dynerf_scenes}, Table~\ref{tab:meetroom_scenes}, and Table~\ref{tab:panoptic_scenes} present comprehensive comparative results between ReCon-GS and recent state-of-the-art methods across three benchmark datasets under specific scene.  

\begin{figure}[htbp]
    \centering
    % 第一行两个图
    \begin{subfigure}[b]{0.48\textwidth}
        \includegraphics[width=\textwidth]{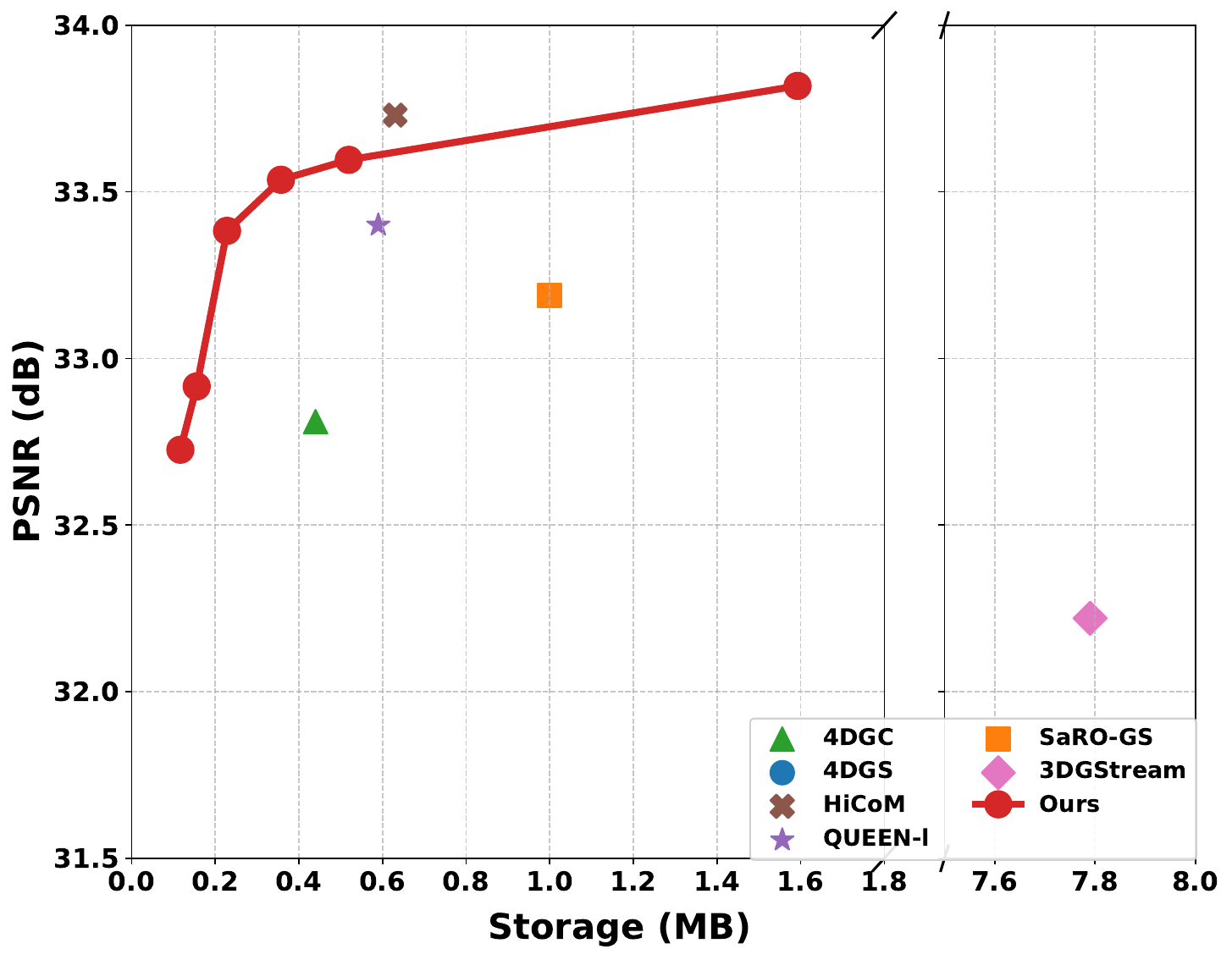}
        \caption{\textit{Cook Spinach}}
        \label{fig:r-d:cook_spinach}
    \end{subfigure}
    \hfill % 在图之间添加一些水平空间
    \begin{subfigure}[b]{0.48\textwidth}
        \includegraphics[width=\textwidth]{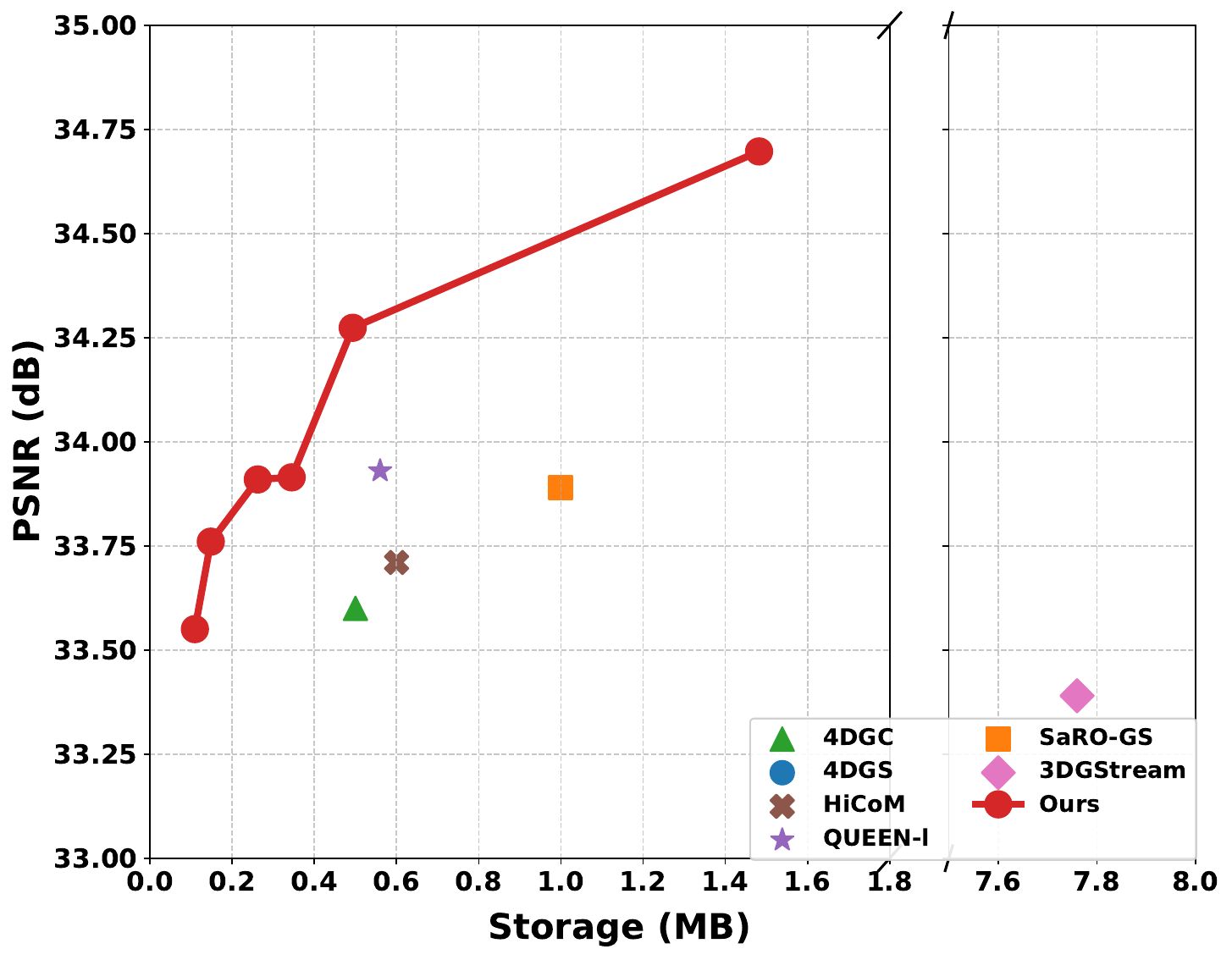}
        \caption{\textit{Sear Steak}}
        \label{fig:r-d:sear_steak}
    \end{subfigure}
    \vspace{5pt}
    % 第二行两个图
    \begin{subfigure}[b]{0.48\textwidth}
        \includegraphics[width=\textwidth]{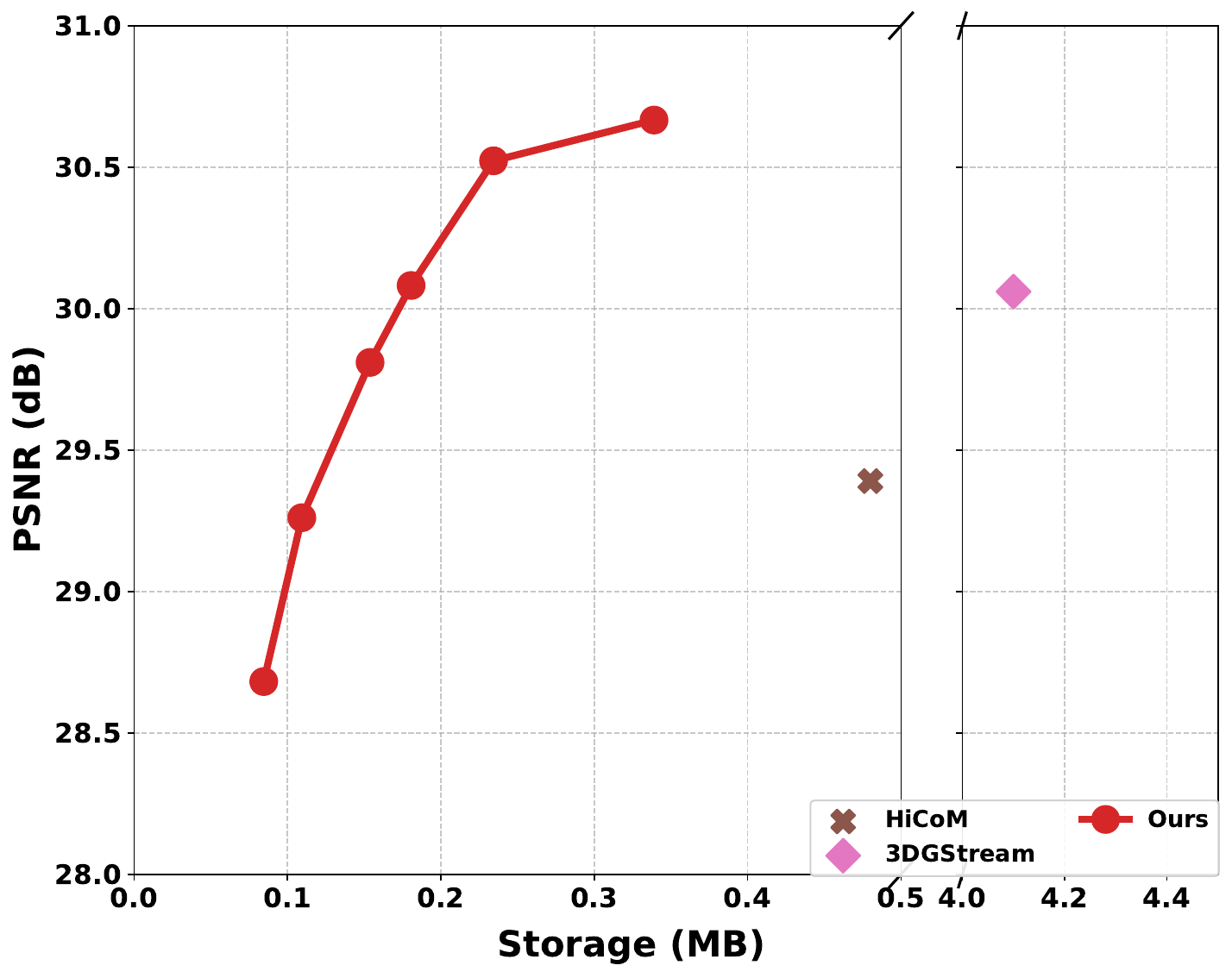}
        \caption{\textit{Discussion}}
        \label{fig:r-d:discussion}
    \end{subfigure}
    \hfill % 在图之间添加一些水平空间
    \begin{subfigure}[b]{0.48\textwidth}
        \includegraphics[width=\textwidth]{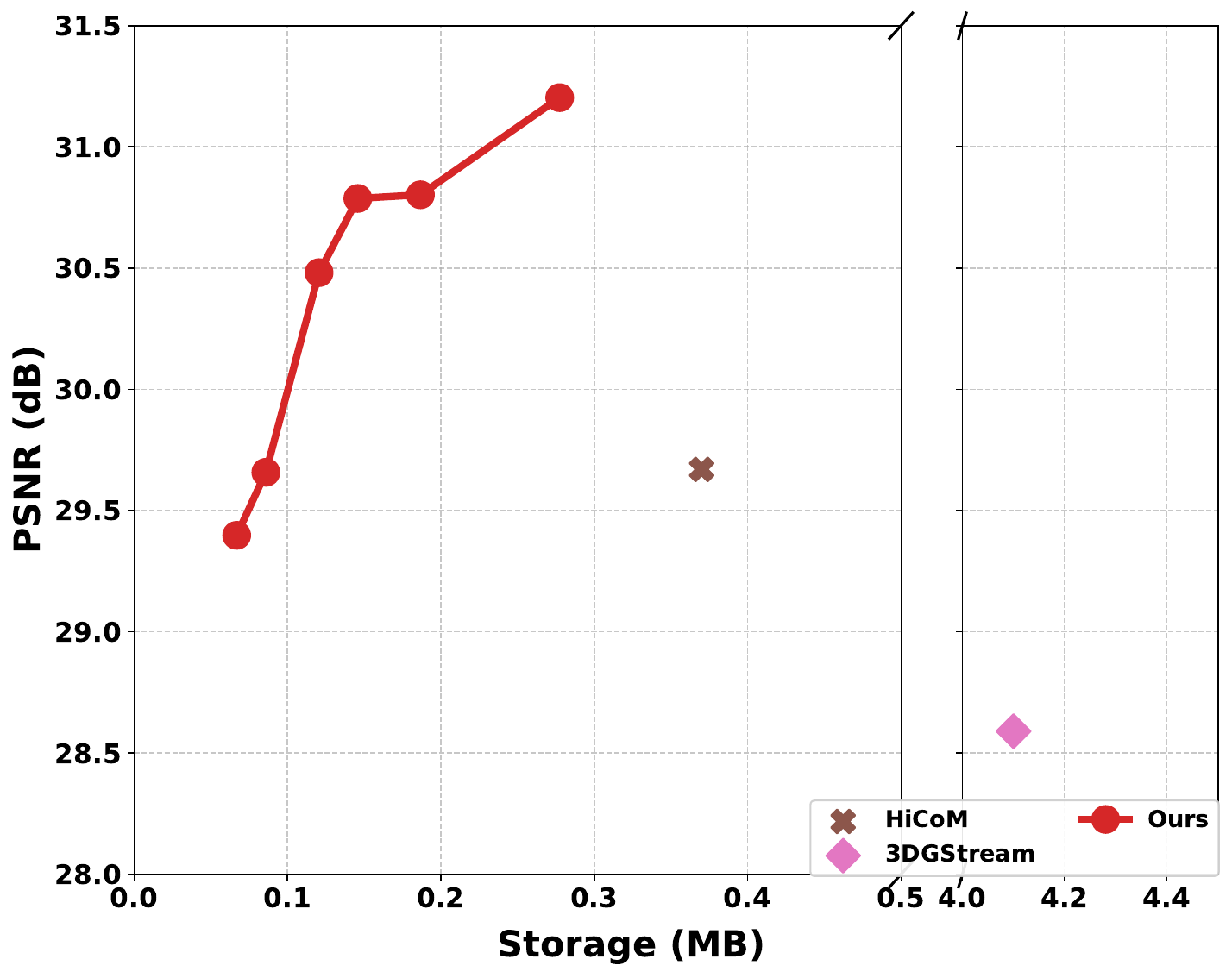}
        \caption{\textit{Trimming}}
        \label{fig:r-d:trimming}
    \end{subfigure}
    
    \caption{ \textbf{Rate-distortion performance across diverse scenarios.} As evidenced by comparative evaluations, our framework not only achieves superior compression efficiency compared to state-of-the-art approachess~\cite{4d-gs,3dgstream,hicom,queen,sarogs,4dgc}, but also demonstrates unique adaptability in storage-aware fidelity optimization - a critical capability for practical deployment scenarios where resource constraints and reconstruction quality must be jointly optimized.}
    \label{fig:r-d_plot}
\end{figure}
\begin{table}[!t]
\centering
\caption{Quantitative Results under different hierarchical anchor density $\{ N^{(l)}_{anchor}\}^3_{l=1}$.}
\vspace{4pt}
\label{tab:anchor_density_results}
\setlength{\tabcolsep}{12pt}
\begin{tabular}{lcccccc}
\toprule
 & \multicolumn{6}{c}{Hierarchical Anchor Density} \\
\cmidrule(lr){2-7}
 & 1/16 & 1/24 & 1/32 & 1/48 & 1/64 & 1/96 \\
\midrule
Storage (MB/Frame) & 0.63 & 0.44 & 0.37 & 0.28 & 0.19 & 0.14 \\
PSNR (dB) & 32.72 & 32.66 & 32.50 & 32.39 & 32.31 & 32.11 \\
\bottomrule
\end{tabular}
\end{table}

\vspace{-4pt}
\subsubsection{More Results}
\vspace{-4pt}
% 此外，图X展示了我们的方法相较于现有的SOTA方法在几个场景下优异的R-D性能。通过调节层级锚点密度并得益于我们的运动表达范式的高效性，ReCon-GS在不同存储需求下都有很好的渲染质量。如图X所示，与最近的SOTA方法相比，ReCon-GS 有效地缓解了时间性能衰减。ReCon-GS通过动态重构策略，有效的避免了空间几何随着时间变形导致锚点高斯的运动表达能力下降的问题。
Figure~\ref{fig:r-d_plot} demonstrates the superior rate-distortion performance of our method compared to existing SOTA approaches across multiple scenarios. Table~\ref{tab:anchor_density_results} presents the detailed results under different hierarchical anchor densities. By modulating hierarchical anchor density $\{ N^{(l)}_{anchor}\}^3_{l=1}$ and leveraging the high-efficiency characteristics of the Adaptively Hierarchical Motion Presentation paradigm, ReCon-GS maintains robust rendering quality under varying storage constraints. 

\begin{figure}[p]
  \centering
  \includegraphics[width=1.0\textwidth]{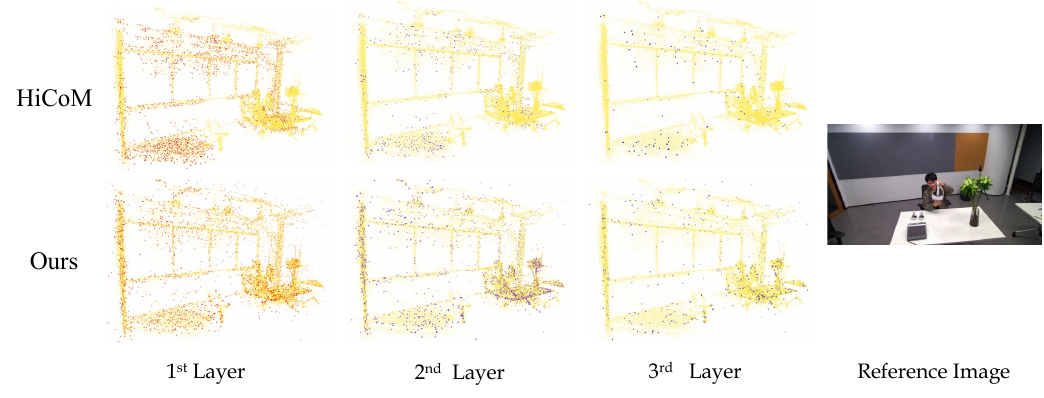}
  \caption{\textbf{The Visualization of Anchor Gaussian Distribution.} $1^{st}$ Layer is the finest layer, and $3^{rd}$ Layer is the coarsest layer. The yellow points represent the General Gaussian, while the red, purple, and blue points represent the Anchor Gaussians for their respective layers.}
  \label{fig:anchor-visualization}
  \vspace{-8pt}
\end{figure}
\vspace{-4pt}
\subsection{Anchor Gaussian Visualization}
\vspace{-4pt}
\label{appendix:anchor_gaus_vis}
% 我们对ReCon-GS和HiCoM锚点高斯在不同层级的分布进行了可视化，结果在图X中展示。从图X中，我们可以清晰的看出，相较于基线方法HiCoM，ReCon-GS将更多表征运动的锚点高斯分配到了运动区域与纹理丰富的区域（人、VR眼睛、花），这主要归因于ReCon-GS使用了基于密度的锚点高斯分配机制。如此一来，ReCon-GS可以对运动物体和纹理丰富的区域进行更加细致的刚体运动划分与表征，从而提升整体的渲染质量。
We visualize the distribution of Anchor Gaussians at different levels for ReCon-GS and HiCoM, with the results shown in Figure~\ref{fig:anchor-visualization}. As shown in the Figure~\ref{fig:anchor-visualization}, compared to the baseline method HiCoM, ReCon-GS assigns more Anchor Gaussians to motion-rich and texture-rich regions ($e.g.$, people, VR glasses, flowers). This is primarily due to the density-based Anchor Gaussian assignment mechanism used by ReCon-GS. As a result, ReCon-GS is able to perform more refined rigid motion representation in areas with rich textures, thus improving the overall rendering quality.

\vspace{-4pt}
\subsection{Qualitative Results}
\begin{figure}
  \centering
  \includegraphics[width=0.9\textwidth]{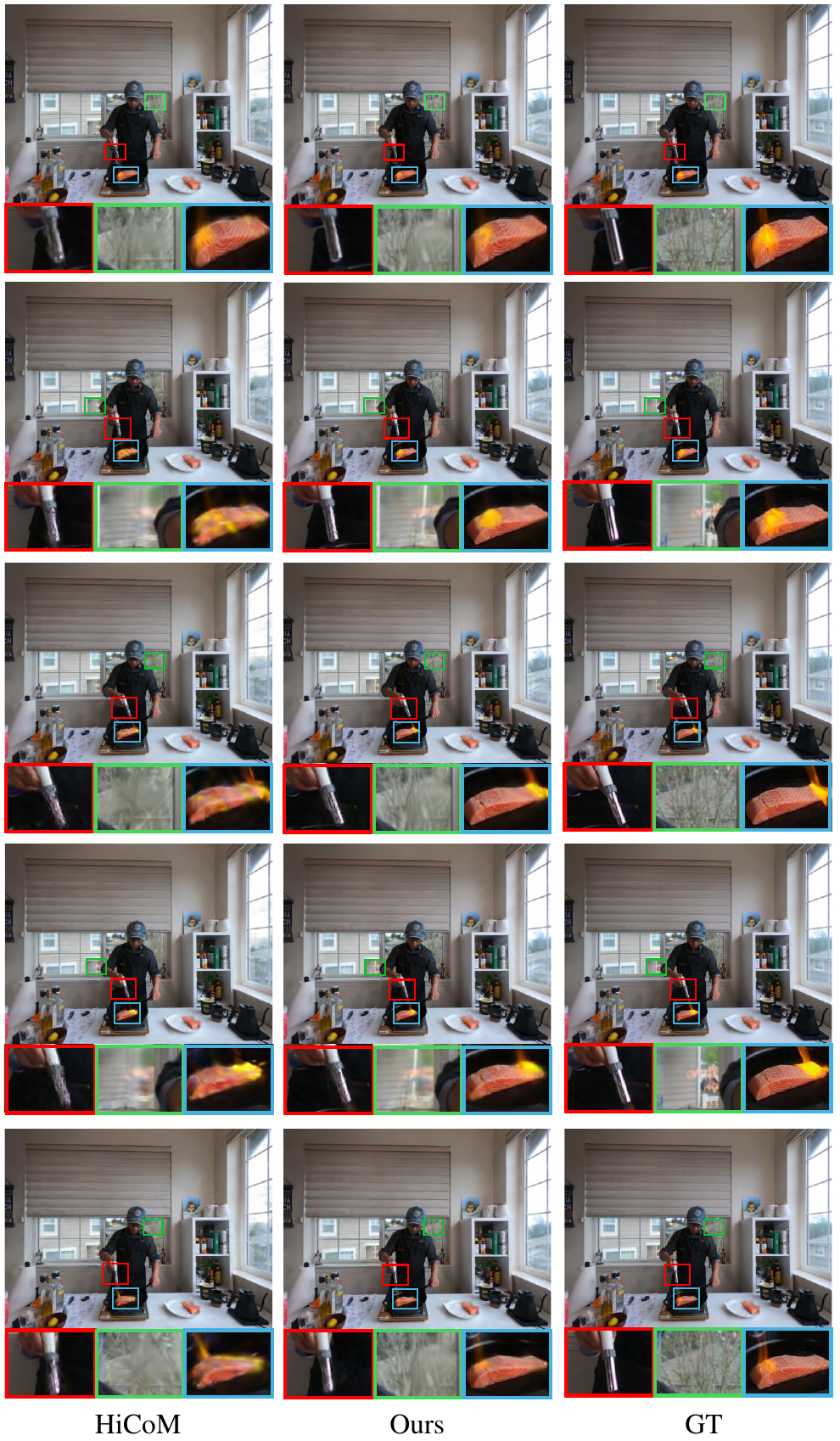}
  \caption{Qualitative results on N3DV Dataset.}
  \label{fig:quality_n3dv}
\end{figure}

\vspace{-4pt}
\label{appendix:qualitative}
% 图X展示了我们的方法与HiCoM以及3DGStream在N3DV数据集更多的数据比较结果。
Figure~\ref{fig:quality_n3dv} provides expanded quantitative comparisons between ReCon-GS and HiCoM on the N3DV dataset, highlighting critical performance differentials in dynamic scene reconstruction.

\begin{figure}
  \centering
  \includegraphics[width=0.9\textwidth]{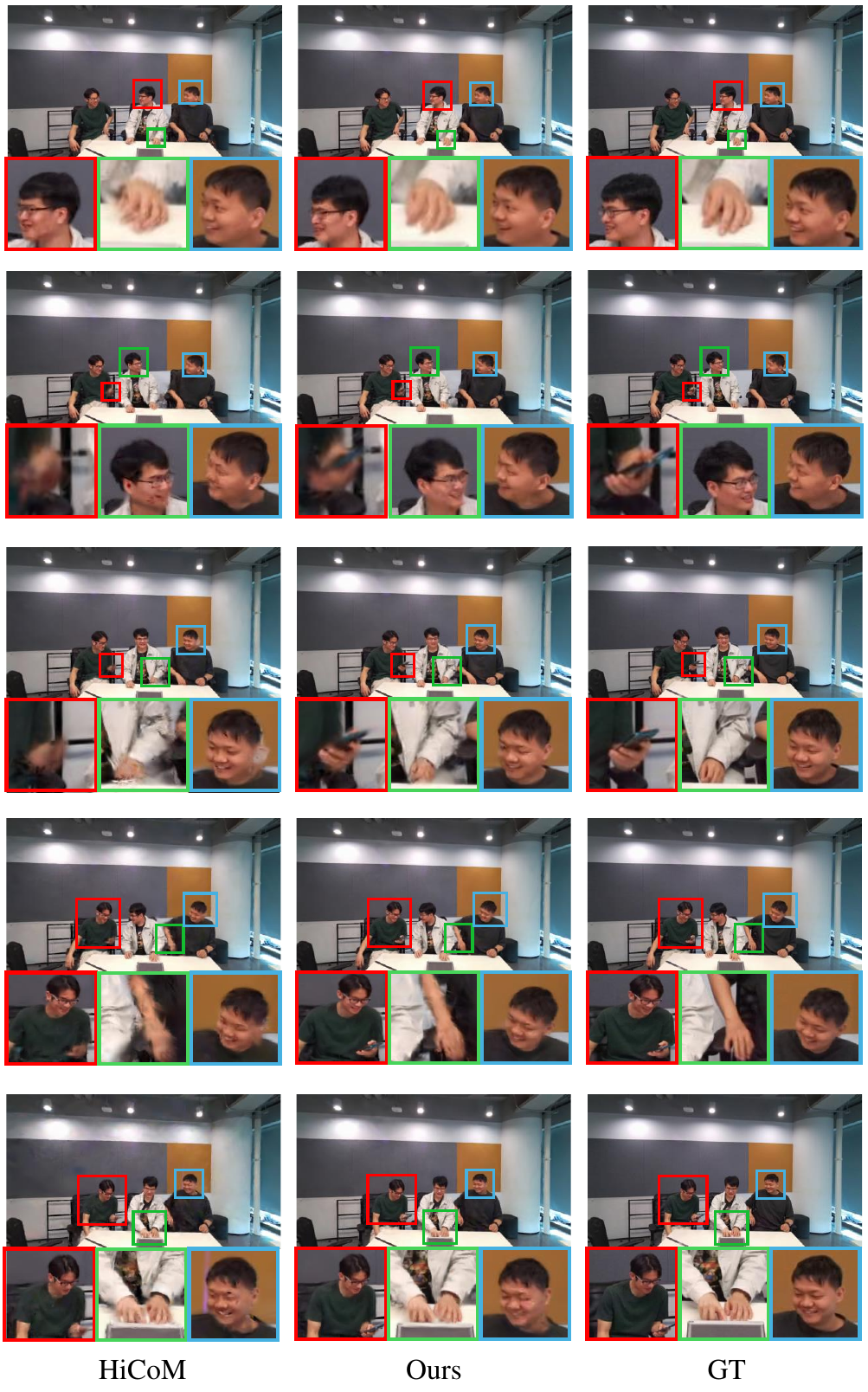}
  \caption{Qualitative results on Meet Room Dataset.}
  \label{fig:quality_meetroom}
\end{figure}

% 图X展示了在MeetRoom数据集上更多的视觉对比结果。
Figure~\ref{fig:quality_meetroom} presents extended visual comparisons on the MeetRoom dataset, contrasting geometric fidelity and temporal consistency across competing frameworks.

\begin{figure}
  \centering
  \includegraphics[width=0.9\textwidth]{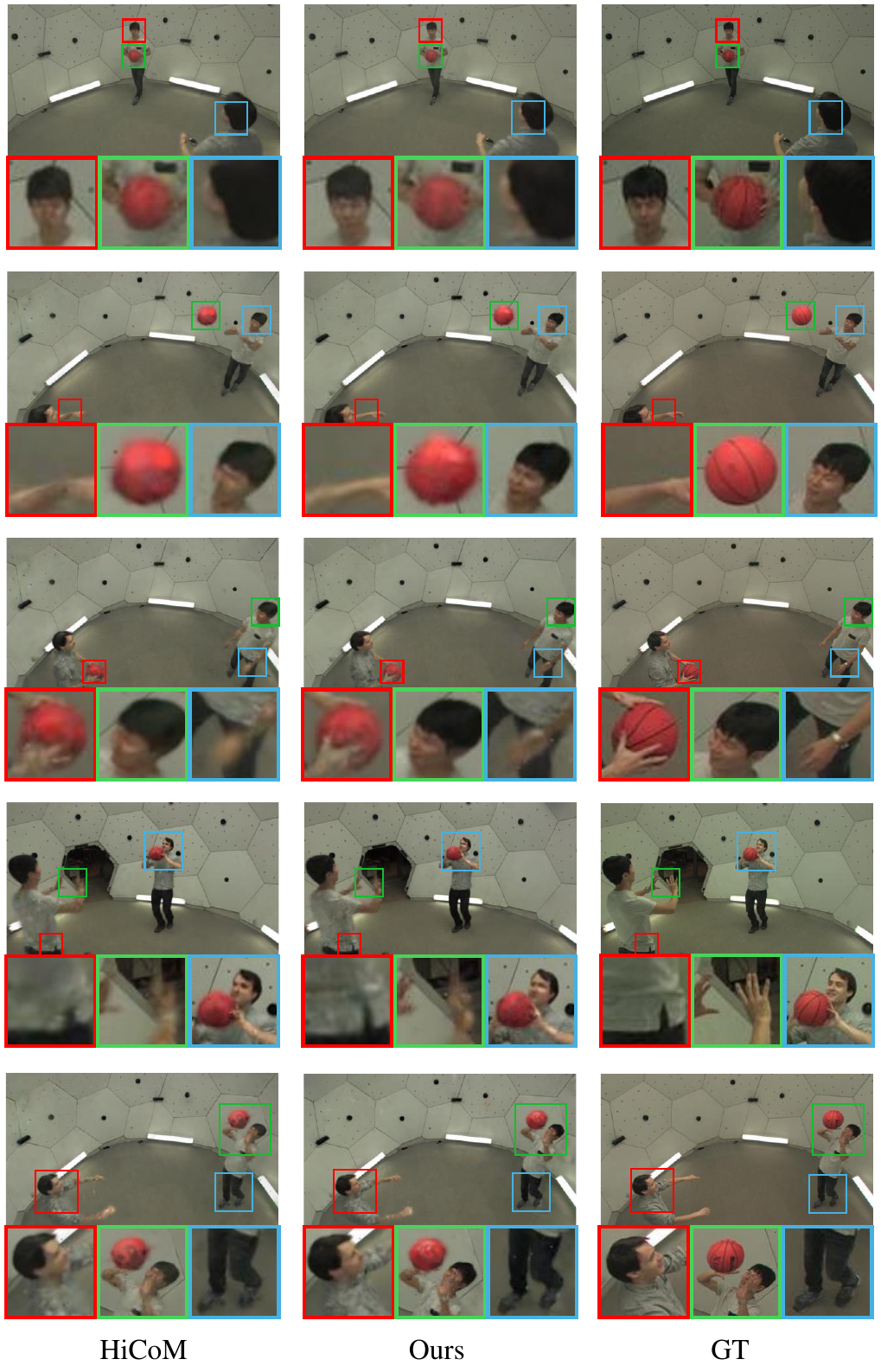}
  \caption{Qualitative results on PanopticSports Dataset.}
  \label{fig:quality_panoptic}
\end{figure}

% 图X展示了在PanopticSports数据集上更多的视觉对比结果。
Figure~\ref{fig:quality_panoptic} details supplementary visual evaluations on the PanopticSports dataset, emphasizing our method's enhanced capability in handling complex motion patterns under real-world capture conditions.

\end{document}